\documentclass[a4paper,11pt]{article}
\pdfoutput=1 
\usepackage{subcaption} 

\usepackage{jcappub} 

\renewcommand*{\vec}[1]{\boldsymbol{#1}}
\newcommand*{\unitvec}[1]{\hat{\vec{#1}}}
\newcommand{\dderiv}{\mathrm{d}}

\newcommand*{\doflat}[1]{\ensuremath{#1^{\mathrm{flat}}}}
\newcommand*{\doearly}[1]{\ensuremath{#1^{\mathrm{early}}}}
\newcommand*{\dolate}[1]{\ensuremath{#1^{\mathrm{late}}}}

\newcommand{\Gearly}{\doearly{G}}
\newcommand{\Glate}{\dolate{G}}
\newcommand{\Aearly}{\doearly{A}}
\newcommand{\Alate}{\dolate{A}}
\newcommand{\field}{\phi}
\newcommand{\fieldflat}{\doflat{\field}}
\newcommand{\fieldearly}{\doearly{\field}}
\newcommand{\fieldlate}{\dolate{\field}}

\newcommand*{\sci}[1]{\times 10^{#1}}
\newcommand*{\unit}[1]{\,\mathrm{#1}}

\title{Scalar and vector tail radiation from the interior of the lightcone}

\author[a]{Craig J. Copi}
\author[a]{Klaountia Pasmatsiou}
\author[a,b]{Glenn D. Starkman}

\affiliation[a]{Physics Department/CERCA/ISO \\
Case Western Reserve University Cleveland, Ohio 44106-7079, USA}
\affiliation[b]{Astronomy Department \\
Case Western Reserve University Cleveland, Ohio 44106-7079, USA}

\emailAdd{cjc5@case.edu}
\emailAdd{kxp265@case.edu}
\emailAdd{glenn.starkman@case.edu}

\abstract{
In a generic spacetime a massless field propagates not just on the surface of the forward lightcone of a source, but in its interior.
This inside-the-lightcone ``tail radiation'' is often described as having ``scattered'' off the spacetime curvature.
In this work, we study the propagation of such tail radiation for a compact, static, spherically symmetric weak-field (i.e.\ low density) mass distribution that is well off the line-of-sight (LOS) between a source and an observer, and that is coupled to the radiation only gravitationally.
For such perturbers, there are four distinct epochs in the observed radiation: the light-cone piece; the subsequent early-time tail --- ending at the first time that a signal moving at the speed of light could travel from the source to a point in the perturber thence to the observer; the subsequent middle-time tail; and the late-time tail, beginning at the last time that a signal could make such a journey.
For massless scalar and vector (eg.\ electromagnetic radiation), we revisit the previously studied early and late-time tail, and perform the first full examination of the middle-time tail.
Studying shorter wavelengths and generic perturbers well off the LOS, we find that the late-time tail carries a small fraction of the energy received by the observer;
however, the total middle-time tail contains much more energy.
We also note that whereas the middle-time tail appears to the observer to emanate from the perturber --- as one might expect for radiation ``scattered'' from the gravitational perturbation --- the late-time tail appears to come back from the source.
We speculate on the potential utility of this middle-time tail for detecting or probing a wide variety of perturbations to the spacetime geometry including  dark matter candidates and dark matter halos.}

\arxivnumber{2008.13069}

\begin{document}
\maketitle
\flushbottom

\section{Introduction}

Signals carried by massless particles are commonly believed to travel through our universe on the lightcone.
This is approximately true of photons (massless vectors) and their associated electromagnetic waves, and gravitons (massless tensors) or at least the associated gravitational waves, and would be approximately true of massless scalars if they existed, because in 3+1-dimensional Minkowski space, the Greens functions for the massless scalar, vector, and tensor wave operators have support only on the null cone.
Of course it is well known that the paths of those particles and their associated waves may be ``bent'' (relative to some hypothetical background homogeneous spacetime) by stress-energy-density inhomogeneities ``nearby'' the null geodesic between the time and place of their emission and the worldline of some observer.
We call this gravitational lensing, and its observation with the correct magnitude was one of the early pieces of evidence supporting General Relativity (GR) \cite{Dyson:1920cwa}.
There may even be multiple null geodesics connecting an emission event and the world-line of an observer --- a phenomena known as strong lensing \cite{10.1111/j.1468-4004.2007.48439.x}.
GR also teaches us that signals may experience time delays when propagating in and out of the gravitational potential wells along their null-geodesic paths from emission to observation \cite{PhysRevLett.13.789}.

It has long been known that a massless field which satisfies a hyperbolic differential equation (such as the wave equation) propagates not only along its characteristic surfaces (the lightcone) but also on the interior of these surfaces \cite{hadamard1923lectures}.
Despite this, in the physics community it is far less commonly known and explored that the Green's functions of the scalar, vector, and tensor wave operators in generic (inhomogeneous, not conformally flat) spacetimes have support both on and inside the null cone.

As first pointed out by Hadamard \cite{hadamard1923lectures}, and succinctly stated by DeWitt and Brehme \cite{DewittBrehme1960}, ``a sharp pulse of light, when propagating in a curved 4-dimensional hyperbolic Riemannian manifold, does not, in general, remain a sharp pulse, but gradually develops a `tail'.''
This is because, as they  further remarked \cite{DewittBrehme1960},
``although [the Green's function] has the same delta-function singularity on the lightcone as in the case of flat space-time, it does not generally vanish inside the lightcone.''
DeWitt and DeWitt \cite{DeWitt:1964aa}, studying the ``radiative damping force on an electrically charged particle falling freely in a static weak gravitational field,'' observed that the particle therefore does not undergo geodesic motion, even if the electromagnetic field vanishes.
There is a non-local term that arises from the non-vanishing of the retarded Green's function inside the lightcone.
They described this physically as arising ``from a back-scatter process in which the Coulomb field of the particle, as it sweeps over the `bumps' in space-time, receives `jolts' which are propagated back to the particle.''
This motivates the description of the non-geodesic terms in the equation of motion of the charged particle as a ``self-force''.
Building on this work that brought the effect to the notice of physicists, Thorne and Kovacs \cite{thorne} derived a formalism for calculating the gravitational waves emitted by any system with weak internal gravitational fields, expressing the radiation in terms of a retarded Green’s function for a weakly curved spacetime, and incorporating among other things what they called ``tail radiation''.

Since then, the tail has gone through phases of renewed interest in the community.
Through the years it has been studied in the context of scalar, vector, and tensor radiation, as well as applied to the self-force problem (see \cite{1993CQGra..10.2699B,2000PhRvD..62h4034M,2001PhRvD..63f3003M}, and references therein for a brief history).
Of direct interest for the work presented here, several authors have employed the Green's function approach in exploring the tail using a variety of techniques to calculate the Green's functions \cite{1995PhRvL..74.2414C,MinoSasakiTanaka1997,1997CQGra..14.1295N,QuinnWald1997,Pfenning_2002,Poisson2011,Chu:2011aa,2013PhRvD..88h4059H,Chu:2020aa}.
We build on this work, in particular the perturbative method developed by Chu and Starkman \cite{Chu:2011aa}.

Wave propagation in curved spacetime using Green's function methods has been studied in a more general context.
A full survey is beyond the needs for the work reported here.
The main focus of the prior work, as related to tail radiation, has been on the late-time tail (see for example \cite{2001PhRvD..63f3003M,2002PhRvD..66d4008P}) --- the radiation arriving at an observer after sufficient time for a signal to propagate at the speed of light from the source to a localized mass concentration (eg.\ a black hole) and then to the observer.
It has been found that this late-time tail radiation can be significant, incorporating even up to twenty percent of the total electromagnetic radiation intensity and fifty percent of the gravitational wave intensity, for very long wavelength radiation \cite{2001PhRvD..63f3003M, 2002CQGra..19..953K} and very specific source-perturber-observer geometries,
though the viability of observing such radiation is uncertain.

Although much work has focused on the tail radiation in black-hole spacetimes, many perturbers of potential interest are not black holes --- exterior black-hole spacetimes may be appropriate outside the perturber, but there is the question of which interior geometry to impose.
Meanwhile, in any weak-field approach, radiation trajectories cannot be studied  arbitrarily close to a point-mass perturber.
This fact is well known and some attempts have been made to avoid doing so \cite{2001PhRvD..63f3003M,Pfenning_2002}.
However, no study has been made of the contribution to the observable signal due to the region interior to the mass distribution.
This is the principal focus of this work, and we will find it leads to surprising results and potentially exciting possibilities.

We begin to explore the potential for direct observation of tail radiation in astrophysical and cosmological settings.
In this first detailed work, we consider the effects on massless scalar and vector (electromagnetic) radiation of a compact, static, spherically symmetric,  gravitational perturber of a Minkowski background.
Specifically, we determine the signal that an observer would detect far from a finite-duration source of massless scalar or vector radiation.
We use the perturbed retarded Green's function in $3+1$ dimensions and we limit ourselves to an everywhere-weakly curved spacetime.
We also limit ourselves to perturbers that couple to the radiation only gravitationally.
This is a preliminary set of calculations that begins the process of probing parameters that may lead to the direct detection and application of tail radiation.

The main new results in this work are the directionality of the tail radiation (where it appears to come from on the sky) and to demonstrate, in detail, the importance of transparent mass distributions.
We will find that the period during which the tail radiation probes the interior of the mass distribution (the middle-time tail) has a greatly enhanced signal compared to earlier or later times, potentially making it detectable.
Our long term goal will be to use this as a starting point to explore questions related to cosmology and astrophysics.
Tail radiation, if detectable in practice, would give us an entirely new window on the universe.
Every inhomogeneity in the metric would be a potential contributor to the signal, and, through the Einstein field equations, so too would every inhomogeneity in the distribution of stress-energy.
This would include both dark matter and ``ordinary'' matter, concentrated or diffuse mass concentrations, overdensities (eg.\ galactic or cluster halos) and underdensities (voids), dark energy if it clumps, gravitational waves however sourced; it could also include more exotic and speculative departures of the metric from flat Friedman Robertson Walker, on whatever scale.
Perhaps we may eventually map the geometry of the universe with these tails, as Eratosthenes measured the geometry of the Earth with the shadow of a stick, long ago.
Furthermore, this will provide additional intuition in exploring black hole questions in the strong field limit where some of the past work can be found in \cite{1974JMP....15..885P, 1993CQGra..10.2699B, 2000PhRvD..62h4034M, 2002CQGra..19..953K, 2003PhRvD..67f4024K}.

We stress that this work is not an exhaustive search of parameters of actual astrophysical systems.
Instead, we present a first exploration of the dependence of the signal on certain physical properties of the system --- the source-perturber-observer geometry, the density distribution of the perturber, the nature of the signal, all done in the weak-gravitational-field limit.
Necessarily, this means that we consider finite-size perturbers, since near enough to any sufficiently small perturber, the weak-field limit is violated.
Others have studied the consequences of deviations from sphericity (i.e.\ higher multipole moments for the gravitational field) and concluded \cite{2002PhRvD..66d4008P} that the monopole moment is the most consequential, though that and the effects of non-static perturbers certainly merit future attention given the properties of realistic mass distributions.

Even with the geometric and weak-field restrictions described above, this system has a multitude of scales:
the Schwarzschild radius of the perturber $r_S=2M$;\footnote{
 Throughout we work in units were $G_N=c=1$.}
 the size of the perturber, $a$;
the impact parameter, $b$, i.e.\ the distance of the perturber from the line of sight (LOS, the line in space joining the source and observer);
 the length of the LOS, $R$;
the duration of a signal pulse, $\tau_p$;
and the time period over which the source is emitting, $\tau_s$.\footnote{
    More precisely we should explore the frequency distribution of the source}
As well, we shall find that the tail radiation depends not just on the mass, $M$, and size, $a$, of the perturber, but also on the profile of the mass distribution (which we characterize simplistically by an additional parameter $p$ in eq.~(\ref{eq:mass_distn})).
The detected radiation, no doubt, also depends on the angular distribution of the signal at the source; however, we shall confine ourselves to the simplest possible sources --- isotropic scalar and dipole vector radiations.

In addition to being constrained by the weak-field limit, $r_S\ll a$, 
we will choose to work in certain limits that enable us to make both analytic and numerical progress ---  $\tau_p\ll a$; $a\ll b$; $\tau_p\ll \tau_s$;  $\tau_s \ll b^2/R $.
For examining the middle-time tail, we will also take the perturber to be equidistant from the source and observer.
These approximations preclude, in particular, any strong gravitational lensing, and for the moment we ignore weak-lensing --- which is a leading-order perturbation of the lightcone signal.
They also prevent us from examining previously studied situations --- such as a source in the geometric shadow of the perturber --- where the late-time tail is known to be able to carry a considerable fraction of the radiated energy that reaches the observer in the long-wavelength limit \cite{2001PhRvD..63f3003M, 2002CQGra..19..953K}.

We are  well aware that there is no known massless scalar field --- certainly none that is sufficiently coupled to ordinary matter for us to have directly observed it.
That this justifies our assumption that the perturbers interact with the scalar radiation only gravitationally --- they have no direct coupling to the massless scalar field --- is cold comfort.
Nevertheless, the scalar case is the most straightforward analytically and numerically, and we present it for its pedagogical value, and for comparison with results for higher spin fields.
It may also prove observationally relevant if axions or axion-like-particles exist.

More problematically, we have made the same assumption that the perturbers have only gravitational interactions with the massless vector radiation, which of course we would like to imagine is electromagnetic radiation.
This is inconsistent with the nature of many known gravitational perturbers, which reflect, absorb, and emit electromagnetic radiation far better than they transmit it; though it may be an appropriate approximation for diffuse objects like globular clusters, galaxies, and galaxy clusters.
Nevertheless, we push ahead to explore parameter values that are more appropriate for non-transparent perturbers in the hopes that the insights we gain are still valuable.

We emphasize our intention in future papers to progressively relax all of these approximations and correct these shortcomings, and more imminently to consider gravitational (i.e.\ massless-spin-2) radiation.
In the gravitational case we will have correctly coupled the radiation to the perturber.

For both scalar and vector radiation, we shall find, as have many before us, that the observer measures no energy or momentum flux in the  ``early-time tail'' --- the period following the arrival of the lightcone signal --- as the retarded Green's function vanishes just inside the lightcone.
However, from the moment when a signal travelling at the speed of light could have propagated from the source to the nearest point on the perturber and then to the observer, a flux of energy and momentum is detectable by the observer.
The resulting tail radiation develops structure reflective of an interplay between the spatial shape of the perturber and the temporal shape of the signal.
We show below that this new `` middle-time tail'' piece appears to come from the direction of the perturber.
It persists until the final time at which the end of the  signal from the source can travel at the speed of light to a point in the perturber, and then to the observer.

The lightcone piece of the signal appears, as is well known, to come from the source.
(Recall that we ignore lensing, so there is no ambiguity in that statement.) 
The  middle-time tail appears,  as one might expect, to come from the perturber.
That it should do so is reassuring but not trivial, since it is sourced not directly by the energy density of the perturber, but by its gravitational potential.
Previous studies of tail radiation have at-best acknowledged the existence of this middle-time tail, and have focused on the subsequent late-time tail; while our investigations suggest the  middle-time tail may be the most interesting part of the tail, and potentially the most observable.

After the  middle-time tail, the signal declines suddenly and precipitously in intensity, and then falls rapidly to zero  in a ``late-time tail''.   
That late-time tail, perhaps surprisingly, appears to the observer to come again from the source not the perturber.

Not unexpectedly, the magnitude of the energy and momentum carried by the tail radiation (both the amount per unit area per unit time, and the time integrals) are small compared to the lightcone piece in all the specific examples we have computed.
However, we also observe that the ratio of the energy-momentum content of the tail to that of the lightcone is a strong function of certain parameters --- such as the impact parameter $b$ --- increasing toward the boundaries of the regions in which our approximations were valid (in this case $b\gg a$).
Recall that it is already known that there are specific situations in which late-time tails  carry large fractions of the detectable very-long-wavelength energy to an observer, but these are unlike generic source-perturber-observer geometries  which are more like those we have considered, and seem to require wavelengths inaccessible within the solar system for electromagnetic radiation.
It is not yet clear whether the detectable tail signals will be dominated by large numbers of typical perturbers, or by rare events.

If observable, tail radiation has potential to become a new window on the universe.
Solar-neutrino tails generated  by Mercury, Venus, the Earth or Moon, could allow us to probe the interiors of those bodies; solar axion tails might be observed through axion-photon mixing during solar eclipses; electromagnetic tails generated by dark matter overdensities or underdensities on all scales could provide information complementary to  lensing about their structure; small tails generated by dark matter candidates, or more exotic physics, might accumulate over cosmological distances and allow us to measure the power spectrum of gravitational perturbations down to otherwise inaccessible length scales, or they could permit us to directly measure the angular velocities of very distant sources; gravitational wave tails could encode the structure of neutron stars and white dwarfs.
None of that is definitively shown to be feasible by the results presented herein, but those results motivate further more exhaustive and intensive investigations.

This manuscript is organized as follows: in section \ref{sec:tail-radiation} we review the formalism of \cite{Chu:2011aa}  and the calculations  of \cite{Chu:2020aa} for the retarded Green's functions of a point-mass perturber and of a compact static spherically symmetric weak-field perturber in a Minkowski background.
In section \ref{sec:scalar-radiation} we apply these calculations to the case of a massless scalar.
In section \ref{sec:vector-radiation} we repeat this exercise for massless vector radiation.
We present a summary of our results in section \ref{sec:conclusions}.

\section{Green's function review}
\label{sec:tail-radiation}

The Green's function method is a classic approach for solving the wave equation.
In this work we follow the approach of Chu and Starkman~\cite{Chu:2011aa} to study the propagation of massless scalar and vector fields by integrating the Green's function in a perturbed spacetime against a source that would produce a known signal in Minkowski spacetime.
The formalism is briefly reviewed here.

In terms of the Green's function in weakly curved spacetime, the field of the scalar radiation is given by
\begin{equation}
  \field(x) = \int  \sqrt{\vert g(x')\vert} \dderiv^{4} x' G(x,x') J(x') ,
  \label{eq:field-scalar}
\end{equation}
and for vector (electromagnetic) radiation the field is given by
\begin{equation}
    F_{\mu \nu} (x) = \int  \sqrt{\vert g(x')\vert}\dderiv^{4} x' \partial_{[\mu} G_{\nu] \beta'} J^{\beta'} (x') ,
    \label{eq:field-vector}
\end{equation}
where $ \vert g(x')\vert$ is the magnitude of the determinant of the background metric at $x'$,
$J(x')$ is the scalar-charge density, $J^{\mu'}(x')$ is the four-vector-current density, and $G(x,x')$ and $G_{\mu \nu'}(x,x')$ are the scalar and vector retarded Green's functions, respectively.

A perturbative approach to calculating these retarded Green's functions  was  derived by Chu and Starkman~\cite{Chu:2011aa} for spacetimes that are close to backgrounds in which the Green's function is known exactly.
Throughout we will use their notation of $[]$ for the antisymmetric product of indices, $\{\}$ for the symmetric product of indices, and the metric signature $(+---)$.
In the case of a background Minkowski spacetime, a static perturbing gravitational potential, and to leading order in the perturbed gravitational field, these Green's functions are given by \cite{Chu:2011aa}\footnote{
    The Green's functions are bitensors so coordinate transformations of $x$ and $x'$ can be applied independently.
    A prime on an index represents which coordinate transformation to use.
    A prime on the index of a derivative means taking derivatives with respect to $x'$.
    See \cite{Chu:2011aa} for a more thorough discussion,}
\begin{align}
  G(x,x') &\approx \frac{\Theta(t-t')}{4 \pi} \left\{ \delta( \sigma_{x,x'}) + \Theta( \sigma_{x,x'}) \hat{I}^{(S)} \right\},
  \label{eq:G-scalar} \\
  \intertext{for a minimally coupled scalar field ($\zeta=0$), and, in the Lorenz gauge,\newline}
  G_{\mu \nu'} (x,x') &\approx \frac{\Theta(t-t')}{4 \pi} \left\{ \delta( \sigma_{x,x'}) g_{\mu \nu'} + \Theta(\sigma_{x,x'})\left[ \eta_{\mu \nu} \hat{I}^{(S)} + \hat{I}_{\mu \nu}^{(A)} + \widehat{(R|1)}_{\mu \nu}\right] \right\}.
  \label{eq:G-vector}
\end{align}
We proceed to explain the various symbols in \eqref{eq:G-scalar} and \eqref{eq:G-vector}.
The first term in curly brackets (with the delta function) is the usual propagation on the lightcone, while the second term (with the step function) is propagation inside the lightcone --- the tail.
Here $g_{\mu\nu'}$ is the parallel propagator between the fixed spacetime points $x$ and $x'$, and $\sigma_{x,x'}$ is Synge's world function, which, in the weak field limit, reduces to
\begin{equation}
    \sigma_{x,x'} \approx \bar{\sigma}_{x,x'} = \frac{1}{2} \left[ (t-t')^2 - |\vec x - \vec x'|^2 \right].
\end{equation}
The parallel propagator $g_{\mu \nu'}$ contains information about the light-cone piece of the propagation from the source to the observer, including gravitational lensing, and gravitational time delay.
Its calculation in this formalism has been described in \cite{Chu:2011aa} for a Kerr metric.
Note that $g_{\mu \nu'}$ encodes important physics along the LOS, while here we are interested in the tail radiation.
Since we insist for this work that the perturber is well enough separated from the LOS that lightcone radiation and tail radiation do not interfere,  weak-field corrections to the light-cone piece may be ignored.
The parallel propagator can therefore be set equal to its Minkowski-space value $\eta_{\mu\nu'}$.
In future work we will want to relax some of these restrictions, and a detailed calculation of $g_{\mu \nu'}$ for a finite-size perturber will be needed.

Inside the lightcone there are several contributions to the Green's functions.
The Ricci tensor to first order in the perturbed metric is
\begin{equation}
    \widehat{(R|1)}_{\beta\nu} = \frac{1}{2} \left( \partial_{\mu^+} \partial_{\{\beta^+}  \hat{I}_{\nu\}}\vphantom{I}^{\mu} -\eta^{\alpha \mu} \partial_{\alpha^+} \partial_{\mu^+} \hat{I}_{\beta \nu} \right),
\end{equation}
while the $\hat{I}$ functions are given by
\begin{align}
\label{Ihat}
  \hat{I}^{(S)} &\equiv {\frac{1}{2}} \partial^{\mu} \partial_{\mu'} \hat{I}  - \partial_{\alpha}  \partial_{\beta'} \hat{I}^{\alpha \beta}, \\
  \hat{I}^{(A)}_{\mu \nu} &\equiv  \frac{1}{2} \partial^{\alpha^-} \partial_{[\mu^+} \hat{I}_{\nu]\alpha},
  \label{eq:antisymtail}
\end{align}
with
\begin{equation}
  \hat{I}_{\alpha \beta} \equiv \delta_{\alpha \beta} A(x,x'),  \qquad \hat{I} = \eta^{\alpha \beta} \hat{I}_{\alpha \beta} = - 2 A(x,x'),
  \label{eq:2Is}
\end{equation}
and $\partial_{\mu^{\pm}}\equiv \partial_{\mu}\pm\partial_{\mu'}$.
Finally, $A(x,x')$ can be calculated as the scattering of the radiation from the gravitational potential of the perturbing mass distribution.

The details of calculating $A(x,x')$ in the first Born approximation can be found in \cite{Pfenning_2002} and  \cite{Chu:2020aa};
we briefly review the results.
We consider a source, an observer, and a static, localized, spherically symmetric mass distribution that weakly perturbs the background  Minkowski spacetime.
The scalar field has been studied before in weakly curved spacetimes using a multipole expansion for a rotating mass distribution where it was found that the static, monopole term describes at least the late-time behavior \cite{2002PhRvD..66d4008P} of the tail radiation, providing justification beyond just the desire for simplicity in choosing a static, spherically symmetric mass distribution in this work.
The reader should note that, in calculating the Green's functions, we have taken the perturbing mass-distribution to be transparent to the radiation under consideration --- i.e.\ coupled only gravitationally.
Whether or not that is appropriate would certainly depend on the specifics of the perturber and the radiation.
We leave the consideration of absorptive or reflective perturbers to future work.

For a source that emits radiation at time $t'$ and a point-mass perturber, it is well-known that\footnote{
    Note that our definition of $A$  is consistent with Pfenning and Poisson~\cite{Pfenning_2002}.
    Comparing~(\ref{eq:Aearly-late}) to Chu and Starkman~\cite{Chu:2011aa} eq.~(106), we see that their $\hat{\mathbb{I}}_{(1)}$ is related to our $A$ by $M\hat{\mathbb{I}}_{(1)}=-A$.}
\begin{equation}
    A_{\mathrm{point}} (x,x') = \begin{cases}
        A_{\mathrm{early}}\equiv - \frac{M}{R} \log \left( \frac{r + r' + R}{r + r' - R} \right), & t-t' < r+r' \\
        A_{\mathrm{late}}\equiv- \frac{M}{R} \log \left( \frac{t - t' + R}{t - t' - R} \right), & t-t' > r+r' ,
    \end{cases}
    \label{eq:Aearly-late}
\end{equation}
where $r\equiv |\vec x|$, $r'\equiv |\vec x'|$, and $R\equiv |\vec x-\vec x'|$.
When $t-t'< r+r'$ this is called the \emph{early-time tail} and when $t-t' > r+r'$ this is called the \emph{late-time tail}.

Examining~(\ref{eq:Aearly-late}) we see that $A_{\mathrm{point}}$ is continuous but not differentiable at the transition from the early-time to the late-time tail, $t-t' = r+r'$.
Furthermore, as $t-t'$ approaches $r+r'$ the radiation probes arbitrarily close to the mass distribution.
This is a problem due to both the usual singularity in the gravitational potential at the location of the point mass and the large value of the potential as the point mass is approached.
The weak-field limit, and with it the first Born approximation, breaks down, and the calculation is invalid.
To circumvent this, a common approach is to ``avoid'' regions too close to the point mass.
We adopt a different approach and instead treat the mass as an extended object that obeys the weak-field limit everywhere.
Intriguingly, we find below that the tail radiation is sensitive to the details of the  mass distribution.

Extended objects were considered in \cite{Chu:2020aa} using the formalism described above and applied to the self-force problem.
In this work we follow the same approach and briefly review it.
The mass distribution is modeled as a sphere of radius $a$ and  density profile
\begin{equation}
  \rho(r\leq a) = {\rho}_0 \left(1-\frac{r^2}{a^2}\right)^p,
  \label{eq:mass_distn}
\end{equation}
with $\rho=0$ for $r\ge a$.
Here $\rho_0$ is determined so that the total mass is some $M$, while $p$ is chosen to ensure that this distribution is sufficiently smooth at the boundary, $r=a$, to allow for the calculation of all our desired quantities (as discussed below).
We have found that $p=2$ is sometimes sufficient, and $p=4$ is always sufficient, for the sources we consider and for both scalar and vector radiation.
We use $p=4$ for the balance of the reported results.
We have not explored the significance of the need for such high levels of differentiability of $\rho(r)$ at the boundary nor by its dependence on the time dependence of the source and the nature of the radiation

\begin{figure}
    \centering
    \begin{subfigure}[b]{0.475\textwidth}
        \centering
        \includegraphics[width=\textwidth]{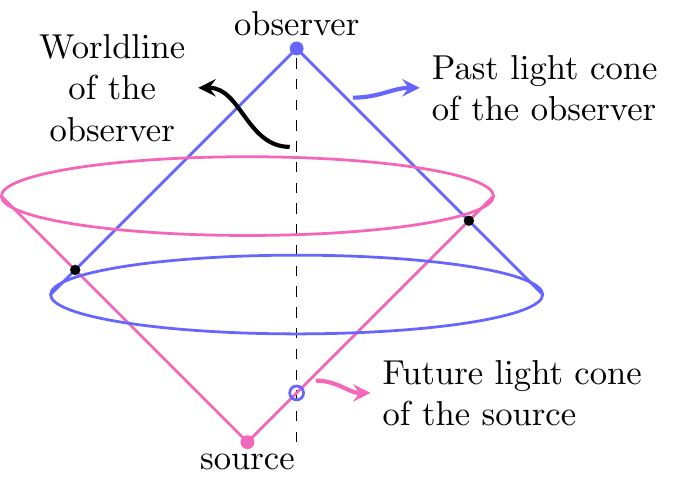}
        \caption{Spacetime diagram of a source event and observer.}
        \label{fig:lightcones}
    \end{subfigure}
    \hfill
    \begin{subfigure}[b]{0.475\textwidth}
        \centering
        \includegraphics[width=\textwidth]{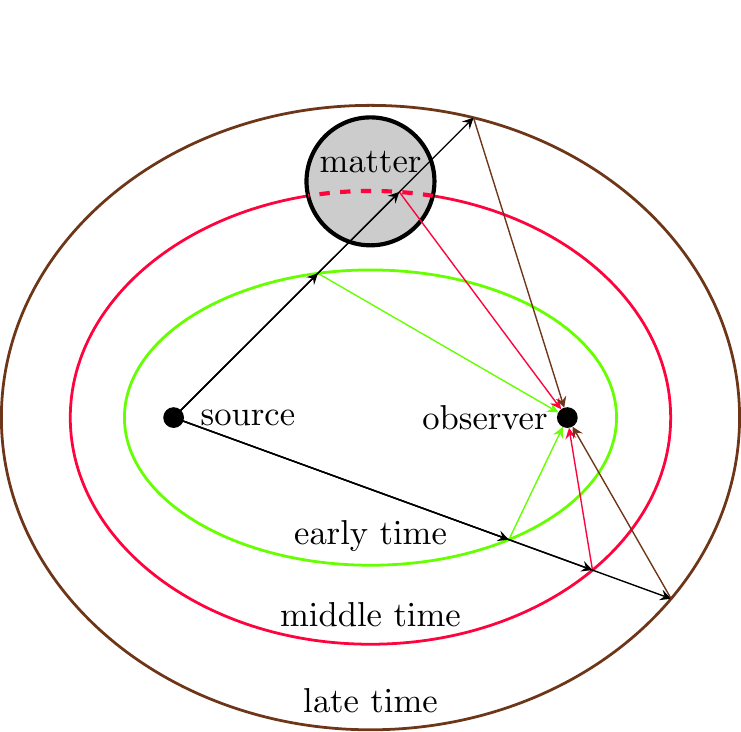}
        \caption{Constant time slice projections onto two spatial dimensions of the future and past lightcone constant time intersections.}
        \label{fig:ellipsoids}
    \end{subfigure}
    \vskip\baselineskip
    \caption{Geometric representation of the tail radiation.
    In the left panel the null signal of a source event is detected by an observer where the observer's worldline intersects the edge of the future lightcone of the source (blue circle).
    At any time after this, the past lightcone of the observer (downwards, blue cone) intersects the future lightcone of the source (upwards, magenta cone).
    Null trajectories from the source may scatter from points in spacetime and arrive at the observer.
    The two small, black disks represent two such locations in spacetime that will allow scattered signals to concurrently arrive at the observer.
    The right panel shows a two dimensional projection of the regions of spacetime that have scattered signals arrive concurrently.
    These regions are confocal ellipses with semi-major axes of length $t=t'$.
    Sample paths for early, middle, and late time trajectories are shown.
    See the text for more details.
    }
    \label{fig:geometry}
\end{figure}

To provide a more intuitive understanding of the tail and to aid in understanding the ensuing mathematics, figure \ref{fig:geometry} illustrates the spacetime geometry of the system under study.
The left panel, figure~\ref{fig:lightcones}, shows the source as a spacetime event (i.e.\ well localized spatially and temporally), while the static observer traces out a vertical worldline (in the direction of increasing time).
The future light-cone of the source (magenta cone) intersects the observer worldline at some particular time denoted by the small blue circle.
This represents the spacetime location of the arrival of the null signal from the source to the observer.
In a conformally flat 3+1-dimensional spacetime, such as 3+1 dimensional Minkowski space, this would be the only signal that the observer could detect.
At all future times, the interiors of the past lightcone of the observer (blue cone) and the future lightcone of the source overlap.
In a 1+1 dimensional spacetime the past and future lightcones intersect at two points represented by the small, black disks.
These intersecting lines represent null signals that could propagate from the source, ``scatter'' from these points in spacetime, and then arrive at the observer (again along null trajectories) in coincidence.
In two and three spatial dimensions the paths of coincidence would scatter from a region in spacetime defined by an ellipse or ellipsoid, respectively.
The panel on the right, figure~\ref{fig:ellipsoids}, shows projections of the scattering regions onto two spatial dimensions for three different times.
We see that the ellipses have foci at the source and observer and that they are confocal with semi-major axis which is equal to $t-t'$.
The ellipses (ellipsoids) represent the surfaces in spacetime at which radiation emitted from the source and travelling at the speed of light can ``scatter'' off the geometry of the curved spacetime (in the weak field limit this means scattering off the gravitational potential) and arrive at the observer concurrently.

With the extended mass distribution~(\ref{eq:mass_distn}) figure~\ref{fig:ellipsoids} shows that there are three distinct periods: in the early-time, the perturbing mass lies wholly outside the ellipse; in the late-time, the perturbing mass lies wholly inside the ellipse; and in the middle-time, part of the ellipse passes through the mass.
That middle period begins with a period when the center of the mass is outside the ellipse, which is followed by a period when it is inside the ellipse.
Sample trajectories are show in figure~\ref{fig:ellipsoids} for the three time periods and show geometrically why the surfaces are ellipses (or ellipsoids in three dimensions).
The black rays represent outgoing radiation.
The colored ellipses and rays show where in spacetime the radiation scatters in order to arrive at the observer concurrently.
Notice in particular for the middle time that some portion of the ellipse/ellipsoid is inside the mass distribution.
Since the curvature (gravitational potential) is different here than on the exterior, the scattered radiation behaves differently.
A main focus of this work is to elucidate this behavior.

Calculating $A(x,x')$ in all of these periods we find
\begin{equation}
    A(x,x') = \begin{cases}
        \Aearly & \gamma < \gamma_0-a \\
        \Aearly - A_\text{interior-x-pt-early} + A_\text{interior-x}, & \gamma_0-a < \gamma<\gamma_0 \\
        \Alate - A_\text{interior-x-pt-late} + A_\text{interior-x}, & \gamma_0 < \gamma<\gamma_0+a \\
        \Alate, & \gamma>\gamma_0+a
    \end{cases} .
    \label{eq:A}
\end{equation}
with
\begin{equation}
    \gamma \equiv \sqrt{\frac{(t - t')^2}{4} - \frac{R^2}{4}}, \qquad
    \gamma_0 \equiv \sqrt{\frac{(r + r')^2}{4} - \frac{R^2}{4}}.
\end{equation}
Here $\Aearly$ and $\Alate$ are given in~(\ref{eq:Aearly-late}), while the other elements of~(\ref{eq:A}) have simple forms :
\begin{align}
  A_\text{interior-x-pt-early} &= - \frac{M (a + \gamma - \gamma_0)}{2\sqrt{\gamma \gamma_0 (\gamma\gamma_0+e^2)}}, 
  \label{eq:Ainterior-x-pt-early} \\
  A_\text{interior-x-pt-late} &= -\frac{M(a - \gamma + \gamma_0)}{2 \sqrt{\gamma \gamma_0 (\gamma\gamma_0+e^2)}} ,
  \label{eq:Ainterior-x-pt-late}
\end{align}
with $e \equiv R/2$.
As noted above, for $A_{\text{interior-x}}$ we almost exclusively use $\rho(r)$ in eq.~(\ref{eq:mass_distn}) with $p=4$, for which
\begin{align}
  A_\text{interior-x} &= - \frac{M}{\sqrt{\gamma \gamma_0 (\gamma \gamma_0+e^2)}} \frac{a}{2048}
 \left[
    793
  - 1386 \left(\frac{\gamma-\gamma_0}{a}\right)^2
  + {} \right. \nonumber \\
 &  \qquad
  {}
   + 1155 \left(\frac{\gamma-\gamma_0}{a}\right)^4
  - 934 \left(\frac{\gamma-\gamma_0}{a}\right)^6
  + 495 \left(\frac{\gamma-\gamma_0}{a}\right)^8
  - {} \nonumber \\
 &  \qquad \left.
  {}
  - 154 \left(\frac{\gamma-\gamma_0}{a}\right)^{10}
  + 21 \left(\frac{\gamma-\gamma_0}{a}\right)^{12}
  \right] .
  \label{eq:Ainterior-x}
\end{align}
Note that the ``interior-x'' appearing in the labels of eqs.~(\ref{eq:Ainterior-x-pt-early})--(\ref{eq:Ainterior-x}) indicates that these results are restricted to the case where the mass distribution is equidistant from the source and observer, as mentioned in the introduction.
(See below for a more thorough discussion of the coordinate system.)

With this expression for $A(x,x')$ we can calculate the Green's functions~(\ref{eq:G-scalar}, \ref{eq:G-vector}).
We emphasize that to obtain \eqref{eq:A}--\eqref{eq:Ainterior-x} we have adhered to the weak-field limit. 
We have also demanded that the size of the spherical perturber, $a$, be much less than other length scales in the problem --- specifically than the distance from the perturber to the null line-of-sight between source and observer, and than the distance from the perturber to the source or the observer.
We note that the appearance of $a$ (and the choice of the density profile through $A_\text{interior-x}$) in the form for $A(x,x')$ already indicates an important difference between the light-cone and the tail radiation.
For all LOS that remain outside the mass distribution, the light-cone radiation (including the effects of gravitational lensing) depends only on the mass of the spherically symmetric static perturber to all orders in perturbation theory because of Birkhoff's Theorem.
Meanwhile, the middle-time tail radiation is always sensitive to the interior structure of the mass distribution.

\section{Scalar radiation}
\label{sec:scalar-radiation}

Although there are no known massless scalar fields in nature, they serve as a simple case for understanding the basic properties of the tail radiation, and they (or at least low mass scalar fields like axions) may some day be discovered.
Here we use the scalar case to present the formalism and develop intuition.

The scalar Green's function~(\ref{eq:G-scalar}) reduces to a simple form
\begin{equation}
    G(x,x') = - \frac{1}{2\pi} \partial_{t} \partial_{t'} A(x,x').
\end{equation}
From~(\ref{eq:A}) we immediately see that, as promised, at early-times
\begin{equation}
    \label{eq:Gearly}
    \Gearly(x,x')=0.
\end{equation}
This is another way of seeing a known property of the tail: \emph{in a Minkowski background, with static spherically symmetric perturbations, there is no early-time tail radiation} at leading order in the perturbations.
Next, again from~(\ref{eq:A}), the late-time Green's function can be calculated to find
\begin{equation}
    \Glate(x,x') = -\frac{2M}{\pi} \frac{t-t'}{\left[ (t-t')^2 - |\vec x-\vec x'|^2\right]^2} .
    \label{eq:Glate-scalar}
\end{equation}
Note that $\Glate$ is independent of the position of the perturber, depending only on the separation between the source and observer and the time between source emission and observation.
These results may seem unsurprising --- after all, during the early time, a signal cannot yet have propagated at the speed of light from the source to the perturber and then to the observer, and during the late time, a signal travelling at the speed of light from the source to the perturber and then to the observer would already have passed the observer.
However, $A(x,x')$ is sourced \cite{Chu:2020aa} by the Newtonian potential, $\Phi(x)$, not just the matter density, $\rho(x)$.
Unlike $\rho(x)$, $\Phi(x)$ is non-zero everywhere.
Thus, while the mathematics is straightforward, the physical result may be unexpected, and may, like Birkhoff's theorem, depend on the very specific assumptions that the perturber is static and spherically symmetric.
Unlike Birkhoff's theorem, which is proven true to all orders in perturbation theory, we make no such claim for these properties of the tail.

To calculate the scalar field at the observer we consider a time-dependent monopole charge, $q(t')$, located at a position $\vec x'_c$ with charge density
\begin{equation}
    J(x') = q(t') \delta^{(3)}(\vec x'-\vec x_c').
    \label{eq:J-scalar}
\end{equation}
Combining this with the Green's function,
and replacing $\sqrt{\vert g\vert}$ by $1$, its leading-order term in the weak field expansion,\footnote{
    The first-order term in $\sqrt{\vert g\vert}$ contributes to first-order corrections to the null-cone piece of the field.
    For the limits in which we are working, the null piece and the tail piece have supports that are well separated in spacetime.
    Because we are focused on the properties of the tail radiation, we are only interested in the leading order of the null radiation.
   }
the field~(\ref{eq:field-scalar}) reduces to
\begin{equation}
    \field(x) = - 2 \int \dderiv t' q(t') \partial_{t}\partial_{t'} A(x,t',\vec x'_c).
\end{equation}
Reassuringly, in the flat-spacetime limit, the field becomes
\begin{equation}
    \fieldflat(x) = \frac{1}{4\pi} \frac{q(t-|\vec x-\vec x_c'|)}{|\vec x - \vec x'_c|},
    \label{eq:field-flat-scalar}
\end{equation}
the usual scalar field for a point charge evaluated at the retarded time.

\subsection{Scalar radiation with point perturber}
\label{subsec:scalar-pointperturber}

As a first example, consider a point-mass perturber.
In this case there are only early and late-time contributions.
We have already seen in~(\ref{eq:Gearly}) that $\Gearly$ vanishes, and so automatically $\fieldearly(x)=0$.
The late-time field is given by
\begin{equation}
    \fieldlate(x) = -\frac{2 M}{\pi} \int \dderiv t' q(t') \frac{t-t'}{\left[ (t-t')^2 - |\vec x-\vec x_c'|^2 \right]^2}.
    \label{eq:field-late-scalar}
\end{equation}
To perform the integral over $t'$, we must specify $q(t')$.

\subsubsection{Sources}
\label{subsec:sources}

In this work, we consider sources that produce a short pulse of null radiation.
We therefore specify $q(t')$ that are non-zero only in some time interval $t'\in[-\Delta t',\Delta t']$, where $\Delta t' \ll R$.
Note that this means that the period of the pulse is $\tau_p=2\Delta t'$.
The two forms we study are given below.

\paragraph{Simple Pulse}

A very simple form for a signal that turns on and off smoothly is
\begin{equation}
    q_{\mathrm{simple}}(t') = q_0 \left[ 1 - \left( \frac{t'}{\Delta t'} \right)^2 \right]^2.
    \label{eq:q-simple}
\end{equation}
Notice that this form satisfies $q_{\mathrm{simple}}(-\Delta t') = q_{\mathrm{simple}}(\Delta t') = 0$.

\paragraph{Sine Pulse}

Generically, $q(t')$ can be decomposed in a Fourier series.
As an initial step towards modeling general sources we also consider a sinusoidal form
\begin{equation}
    q_{\mathrm{n}}(t') = \begin{cases}
    q_0 \sin\left( \frac{n \pi t'}{\Delta t'} \right), & \vert t'\vert\leq \Delta t'\\
    0, & \vert t'\vert > \Delta t'
    \end{cases},
    \label{eq:q-sine}
\end{equation}
though we are still restricting the source to a finite time interval, $t'\in[-\Delta t',\Delta t']$.
The pulse frequency can be changed independent of $\Delta t'$,
\begin{equation}
    f_n = \frac{n}{2\Delta t'}.
\end{equation}
Changing $n$ for fixed $\Delta t'$ allows for  multiple oscillations in the time interval.

With these simple smooth sources, we are able to perform the integral over $t'$ in eqs.~\eqref{eq:field-flat-scalar} and \eqref{eq:field-late-scalar}.
The assumption of a short pulse means that we can expand the Green's function in the integrand to leading order in $t'$ prior to performing the integral.
More explicitly, the Green's function is Taylor expanded around $t'=0$ and truncated at order $m$, determined by the smallest value of $m$ satisfying
\begin{equation}
    \int_{-\Delta t'}^{+\Delta t'} \dderiv t' t'^m q(t') \ne 0.
\end{equation}

\paragraph{Field of the simple pulse}

For the simple pulse we find, using a zeroth-order expansion in $t'$,
\begin{align}
    \field_{\mathrm{simple}}^{\mathrm{flat}}(x) &= \hphantom{-} \frac{q_0}{4\pi R} \left[ 1 - \left( \frac{t-R}{\Delta t'} \right)^2 \right]^2, \\
    \field_{\mathrm{simple}}^{\mathrm{late}}(x) &= -\frac{32 q_0}{15 \pi} \frac{M t \Delta t'}{(t^2-R^2)^2}.
\end{align}

\paragraph{Field of the sine pulse}

For the sine pulse we find, using a first-order expansion in $t'$,
\begin{align}
    \fieldflat_n(x) &= \frac{q_0}{4\pi R} \sin\! \left( \frac{n\pi(t-R)}{\Delta t'} \right), \\
    \fieldlate_n(x) &= \frac{4(-1)^n q_0}{n \pi^2} \frac{M \Delta t'^2 (3t^2 + R^2)}{(t^2-R^2)^3}.
\end{align}

\subsubsection{Energy density}

For detecting radiation, we are interested in the energy and momentum it carries.
The energy density of a real scalar field is the time-time component of the associated stress-energy tensor,
\begin{equation}
    T^{00} = \frac{1}{2} \left[ (\partial_t \field)^2 + (\vec\nabla \field)^2 \right].
    \label{eq:T00-scalar}
\end{equation}
For the null signal, it will be convenient to consider the energy density at the observer averaged over the pulse,
\begin{equation}
    \langle T^{00;\mathrm{flat}} \rangle \equiv \frac{1}{2\Delta t'} \int_{R-\Delta t'}^{R+\Delta t'} T^{00}_{\mathrm{flat}} \,\dderiv t .
    \label{eq:T00avg-scalar}
\end{equation}

For the simple pulse, evaluating the derivatives, and averaging  the null signal,
\begin{align}
    \langle T^{00;\mathrm{flat}}_{\mathrm{simple}} \rangle &= \frac{8 q_0^2}{105 \pi^2 \Delta t'^2 R^2}, \\
    T^{00;\mathrm{late}}_{\mathrm{simple}} &= \frac{512 q_0^2 M^2 \Delta t'^2 (9t^4 + 22 t^2 R^2 + R^4)}{225 \pi^2 (t^2-R^2)^6}.
    \label{eq:T00scalarsimple}
\end{align}
Since the normalization of the strength of the source is arbitrary, it is more interesting to compute the ratio of the tail energy density to the usual flat-space null signal:
\begin{align}
    \frac{T^{00;\mathrm{late}}_{\mathrm{simple}}}
    {\langle T^{00;\mathrm{flat}}_{\mathrm{simple}} \rangle}
    &= \frac{448  M^2 \Delta t'^4 R^2 (9t^4 + 22 t^2 R^2 + R^4)}{15 \pi^4 (t^2-R^2)^6}.
\end{align}
Similarly, for the  sine pulse
\begin{align}
    \langle T^{00;\mathrm{flat}}_n \rangle &= \frac{n^2 q_0^2}{32 \Delta t'^2 R^2}, \\
    T^{00;\mathrm{late}}_n &= \frac{128 q_0^2 M^2 \Delta t'^4 (9t^6 + 43 t^4 R^2 +19 t^2 R^4 + R^6)}{n^2 \pi^4 (t^2-R^2)^8},
    \label{eq:T00scalarsine}
\end{align}
and so the ratio now becomes
\begin{equation}
    \frac{T^{00;\mathrm{late}}_n}{\langle T^{00;\mathrm{flat}}_n \rangle} =
    \frac{2^{12}  M^2 \Delta t'^6 R^2 (9t^6 + 43 t^4 R^2 +19 t^2 R^4 + R^6)}{n^4 \pi^4 (t^2-R^2)^8}.
\end{equation}

As is clear from previous studies of the late-time tail, whereas the light-cone signal from a point source  has the same duration to leading order at the observer as at the source, once the (middle-time) tail radiation arrives at a particular inertial observer, tail radiation persists indefinitely.
This is an important to note in comparing the light-cone piece and the tail radiation.

\subsubsection{Radiation at infinity}
\label{sec:scalar-radiation-approximation}

We see that there is energy in the late-time tail radiation, $T^{00}_{\mathrm{late}}\ne 0$.
We can go a step further and calculate the energy from the source radiated to infinity in the tail.
This energy is determined from
\begin{equation}
    \Delta E = \int P \,\dderiv t,
\end{equation}
where the power
\begin{equation}
    P = \int_{S^2_\infty} T_{0\mu} n^\mu R^2 \, \dderiv\Omega.
\end{equation}
The integral in $P$ is over a two-sphere at spatial infinity, with $n^\mu$ a spacelike unit vector normal to the sphere --- in Cartesian coordinates $n^\mu=(0,x^i/R)$.
Thus $T_{0\mu} n^\mu = T_{0 i} x^i/R = T_{0 R}$, with the required components of the stress-energy tensor given by
\begin{equation}
    T^{0 i} = - \frac{\partial\field}{\partial t} \frac{\partial\field}{\partial x^i}.
\end{equation}
For the null radiation, the energy density falls as $1/R^2$, thus energy is radiated arbitrarily far from the source.
For the late-time tail it is less immediately clear if this is the case.

For the far-field radiation,
we are interested in the limit $R\rightarrow\infty$ of the power,
with the position of the source, $\vec x_c'$,  fixed.
To study this limit, we write
\begin{equation}
    R_c \equiv |\vec x-\vec x_c'| = \sqrt{r^2+r_c^{\prime 2}-2rr_c'\cos\psi},
    \label{eq:Rc}
\end{equation}
where $\psi$ is the angle between $\vec x$ and $\vec x_c'$.
For $R$ large, $r$ is also large, and in particular $r\gg r_c'$ so $R\approx R_c\approx r$.
Converting to spherical coordinates $(R,\theta,\varphi)$ centered on the source, this limit also means that
\begin{equation}
    \psi \approx \pi-\theta.
\end{equation}
Further, the transition to the late-time behavior of the tail occurs when
\begin{equation}
    t = r + r_c'.
\end{equation}
Putting these limits together, expanding the fields in  $r_c'/R$, and keeping only the leading-order radiation terms --- that is the ones that fall as $1/R^2$ --- we find for the simple and sine pulse-shapes respectively
\begin{align}
    T^{0R;\mathrm{late}}_\mathrm{simple} &\approx  \frac{256 q_0^2 M^2 \Delta t'^2}{225 \pi^2 r_c^{\prime 6} (1-\cos\theta)^6 R^2},
    \label{eq:scalar-T0R-simple} \\
    T^{0R;\mathrm{late}}_{n} &\approx \frac{36 q_0^2 M^2 \Delta t'^4}{n^2 \pi^4 r_c^{\prime 8} (1-\cos \theta)^8 R^2}.
    \label{eq:scalar-T0R-sine}
\end{align}

These momentum-density components both diverge as $\theta\rightarrow 0$.
This limit corresponds to the mass distribution being very close to the LOS between the center of the monopole source and the observer.
The null photons would then pass very close to or through the mass distribution.
In this limit we would expect lensing and/or finite size effects to be important.
For a point mass, both the weak-field limit and the first Born approximation break down.
This implies a lower limit on $\theta$, only above which we can trust the late-time-tail calculation we have performed.
Let $\ell \ll r'_c$ be a characteristic length scale associated with the perturbing mass distribution.
Our approximate momentum components are valid only for $\theta\gg\epsilon\sim \ell/r_c'$.

We can only calculate the power in the tail radiation in that part of the sphere where our approximations apply,
\begin{equation}
    P^{\mathrm{tail}} \approx \int_{0}^{2\pi} \dderiv\varphi \int_{\ell/r'_c}^\pi \sin\theta \, \dderiv\theta \, R^2 T^{\mathrm{late}}_{0R}.
    \label{eq:power-approximation}
\end{equation}
Since the integral over $T^{0R}$ diverges as $\theta\to 0$,
the leading order contribution  comes from the region of the integral closest to the LOS, and
\begin{align}
    P^{\mathrm{tail}}_{\mathrm{simple}} &\approx -\frac{2^{14} q_0^2 M^2 \Delta t^{\prime 2} r_c^{\prime 4}}{1125 \pi\ell^{10}} , \\
    P^{\mathrm{tail}}_n &\approx -\frac{9216 q_0^2 M^2 \Delta t^{\prime 4} r_c^{\prime 6}}{7 n^2 \pi^3 \ell^{14}} ,
\end{align}
where the negative sign indicates that energy is flowing outward from the source.

To find the total energy that escapes to infinity, we integrate the power over the time the energy is flowing.
In principle for the tail the radiation  persists forever.
However, by inspection of \eqref{eq:T00scalarsimple} and \eqref{eq:T00scalarsine}, we see that the radiation energy density is dominated by the time right after it arrives; the same is true of the energy flow.
Since the period of the source pulse is $2 \Delta t'$ the energy can be approximated as
\begin{equation}
    \Delta E^{\mathrm{tail}} \approx 2 \Delta t' P^{\mathrm{tail}}.
\end{equation}
This gives
\begin{align}
    \Delta E^{\mathrm{tail}}_{\mathrm{simple}} &\approx -\frac{2^{15} q_0^2 M^2 \Delta t^{\prime 3} r_c^{\prime 4}}{1125 \pi \ell^{10}} , \\
    \Delta E^{\mathrm{tail}}_n &\approx -\frac{1}{n^2}\frac{18432 q_0^2 M^2 \Delta t^{\prime 5} r_c^{\prime 6}}{7 \pi^3 \ell^{14}} .
\end{align}
Thus energy escapes from the source to infinity in the tail --- there is real energy carried by the late-time tail.

Looking more closely at the expressions for the energy loss we see that
\begin{align}
    \label{eq:Eloss}
    \Delta E^{\mathrm{tail}}_{\mathrm{simple}} &\sim -\left( \frac{q_0^2}{\ell} \right) \left( \frac{M}{\ell} \right)^2 \left( \frac{\Delta t'}{\ell} \right)^3 \left( \frac{r_c'}{\ell} \right)^4 , \\
     \Delta E^{\mathrm{tail}}_{n} &\sim -\left( \frac{q_0^2}{\ell} \right) \left( \frac{M}{\ell} \right)^2 \left( \frac{\Delta t'}{\ell} \right)^5 \left( \frac{r_c'}{\ell} \right)^6.
\end{align}
To determine whether~(\ref{eq:Eloss}) represents a significant energy loss, we compare to the energy transmitted by the null signal.

\subsubsection{Tail/Null-radiation comparison}

Since the mass distribution is required to be far from the LOS between the source and the observer, we can compare the energy lost in the late-time tail to the energy carried by a null signal in flat spacetime.
In flat spacetime, using the same procedure as above we find
\begin{align}
    \Delta E^{\mathrm{flat}}_{\mathrm{simple}} &\approx - \frac{64 q_0^2}{105\pi \Delta t' }, \\
    \Delta E^{\mathrm{flat}}_n &\approx - \frac{n^2 \pi q_0^2}{4 \Delta t' }.
\end{align}
Putting together with the tail results,
\begin{align}
    \label{eq:tailtonullsimple}
    \frac{\Delta E^{\mathrm{tail}}_{\mathrm{simple}}}{\Delta E^{\mathrm{flat}}_{\mathrm{simple}}} &\approx
    \frac{10752}{225} \left(\frac{M}{\ell}\right)^2 \left(\frac{\Delta t^{\prime} r_c^{\prime}}{ \ell^{2}}\right)^4 ,  \\
    \label{eq:tailtonullsine}
    \frac{\Delta E^{\mathrm{tail}}_{\mathrm{n}}}{\Delta E^{\mathrm{flat}}_{\mathrm{n}}} &\approx
    \frac{73728}{7\pi^4 n^4} \left(\frac{M}{\ell}\right)^2 \left(\frac{\Delta t^{\prime} r_c^{\prime}}{ \ell^{2}}\right)^6 .
\end{align}

These ratios are not obviously small.
In order to remain in the weak-field limit, we need $M/\ell\ll1$.
But $r_c'$ is a cosmological distance, and so it seems like the ratio ${\Delta t^{\prime} r_c^{\prime}}/{ \ell^{2}}$ can be large.
Unfortunately, for our calculations to be valid we require the time delay between the null radiation and the late-time tail to be longer than the duration of the pulse.
In the small $b$ limit, that delay is $2 b^2/r_c'$.
Again, since the mass distribution is far from the LOS, $b>\ell$, so we require ${\Delta t^{\prime} r_c^{\prime}}/{ \ell^{2}}<1$.
Thus, as expected, the tail-to-null ratios \eqref{eq:tailtonullsimple} and \eqref{eq:tailtonullsine} must be small.
A length scale that satisfies these constraints and that will allow for numerical estimation of the ratio is the Einstein radius.
For an observer far from the source and lens the Einstein radius reduces to
\begin{equation}
    \ell = r_E \approx \sqrt{4M r_c'}.
\end{equation}
Borrowing parameters studied in more detail below, for a perturber with $M=20M_\odot$, a distance $r'_c=100\unit{Mpc}$ from the source, and single pulse of radiation with wavelength $\lambda=100\unit{m}$ (which corresponds to $\tau_p=\tau_s\approx 3.33\sci{-7}\unit{s}$) we find
\begin{equation}
    \frac{\Delta E^{\mathrm{tail}}_{\mathrm{simple}}}{\Delta E^{\mathrm{flat}}_{\mathrm{simple}}} \sim 7\times10^{-33}.
    \label{eq:scalar-energy-ratio}
\end{equation}

This is quite small and naively would suggest that the tail is not of observational importance.
However, this calculation is only a lower limit; only part of the power was included in the calculation~(\ref{eq:power-approximation}).
In fact, for a point-mass perturber, the excluded power is divergent, suggesting that the actual power radiated could be quite large.
To study this we shall find below that it is in fact important to resolve the mass distribution, and that the largest contribution to the tail radiation is from the time that the interior of the mass distribution can be probed, the middle-time tail.
It might well be that this ratio $E^{\mathrm{tail}}/E^{\mathrm{flat}}$ is related to geometric concepts that would be relevant for doing the calculation generically for non-transmissive perturbers.
We leave this for future work.

\subsection{Scalar radiation with a transmissive compact perturber}
\label{subsec:scalar-compactperturber}

\begin{figure}
  \centerline{\includegraphics{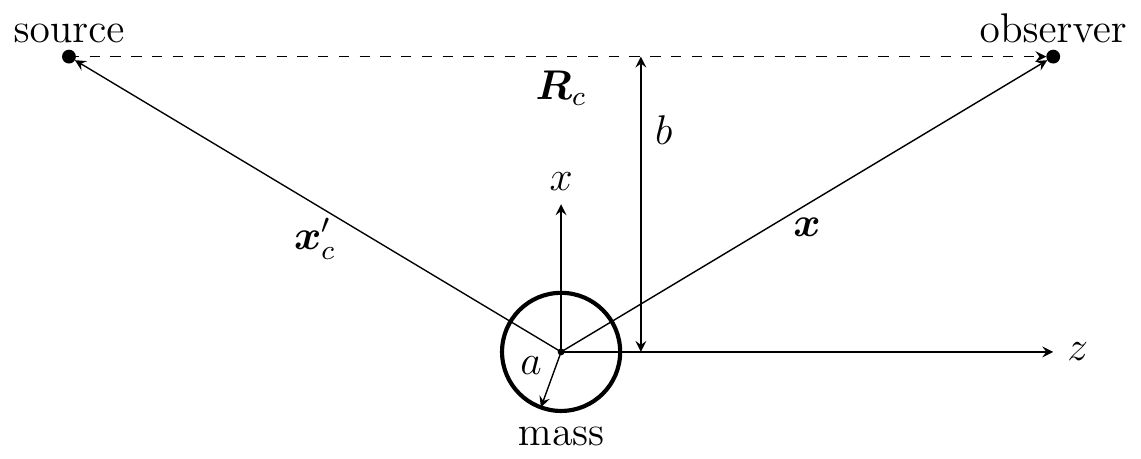}}
  \caption{Coordinate system for a finite-size mass distribution.
  The coordinate system has its origin at the center of the mass distribution.
  The source, observer, and center of the mass distribution lie in the $xz$-plane with the source and observer along a line parallel to the $z$-axis.
  The source and observer are equidistant from the center of the mass distribution so that $|\vec x| = |\vec x'_c|$.
  Not drawn to scale.
  }
  \label{fig:basic-coordinate-system}
\end{figure}

A finite-size mass distribution large enough for the weak-field limit to apply everywhere smooths out the singularity of the point mass and allows for the tail radiation to be studied at all times.
We find below that the ``middle-time tail'' --- the portion of the tail arising from  when the ellipsoid passes through the interior of the mass distribution --- makes a significant contribution to the tail radiation, and indeed dominates the tail radiation.

Introducing a finite-size mass distribution introduces new scales in the problem, and thereby complicates the analytic calculations.
To make progress in this initial study, we make some useful simplifications and approximations.
Most important is that we assume that the mass distribution is transmissive --- it interacts with the radiation only gravitationally.
We study the simplified case where the mass distribution is equidistant from the source and the observer $r=r'$.
We define a coordinate system with its origin at the center of the mass distribution.
Without loss of generality, we  choose the LOS to be parallel to the $z$-axis, and in the $xz$-plane, as shown in figure~\ref{fig:basic-coordinate-system}.
We refer to  $b$, the distance between the LOS and center of the perturber, as the impact parameter.

For all calculations below,
the mass distribution is a sphere of radius $a$, total mass $M$,
and radial mass density $\rho(r)$ given by \eqref{eq:mass_distn}, centered at the origin;
the observer is located at $\vec x=(x,0,z)$;
the center of the source of radiation is located at $\vec x'_c=(x,0,-z)$.
Finally, we require that $r\equiv|\vec x|=r'\equiv|\vec x'_c|\gg a$, $b\gg a$, and the length of the pulse, $\tau_p$, is much shorter than the time delay between the null signal and the arrival of the tail radiation, $\tau$, i.e.\ $\tau_p\ll\tau$.
For a pulse, $\tau_p=2\Delta t'$ and the time delay is $\tau = 2r-2z$.
When the time delay is large (which is equivalent to $r\approx b$), it is easy to satisfy $\Delta t'\ll \tau$.
When $b\ll r$ we have
\begin{equation}
  \tau = 2 r \left( 1 - \sqrt{1-\frac{b^2}{r^2}}\right)
  \approx 2r \left( 1 - 1 + \frac{b^2}{2r^2} \right) = \frac{b^2}{r}.
\end{equation}
Thus, $\Delta t' \ll \tau$ means that
\begin{equation}
  \frac{\sqrt{\Delta t' r}}{b} \ll 1.
\end{equation}
Putting all this together allows us to expand to leading order in small ratios of scales:
$M/a$, $a/b$, and $\sqrt{\Delta t' r}/b$.
\subsubsection{Parameter choices and source properties}

To study the effects of the finite-size mass distribution, we choose  parameters describing the properties of the source, the observer, and the perturbing mass distribution.
These parameters must be chosen to be consistent with all the approximations described above.
For this preliminary work, we are interested in exploring general behaviors, not modeling fully realistic physical systems.
For example, while it is required that the mass distribution interacts with the radiation only through gravitational scattering, we will not confine ourselves to parameter choices where there are known physical realizations.
Our choices of parameters will therefore be motivated by physical systems, but they are not meant to be realistic models of these systems.

Two important properties of the source are its frequency, and its duration.
In this work we consider high frequency (short wavelength) sources as these are relevant for the observation of electromagnetic (vector) radiation.
We reserve the study of low frequency radiation, which is relevant for the observation of gravitational (tensor) radiation, for a follow-on paper.
In the high frequency regime our results will be based on the longest wavelength of electromagnetic radiation that can propagate through the interstellar medium into the solar system, and choose  $\lambda=100\unit{m}$, corresponding to  $f\approx 3\sci{6}\unit{Hz}$.
The frequency dependence of the results will be discussed in more detail below.

For the choice of other parameters we consider two cases based on astrophysical and cosmological scales.

\paragraph{Cosmological Star}

We consider a system based loosely on a massive star at cosmological distances, and will refer to it as the ``cosmological star.''
We choose the perturber mass to be $M=20\unit{M_\odot}$.
We take $a=80\unit{R_\odot}$  --- larger radius than a main-sequence star of that mass, but smaller than a supergiant of that mass.
We fix the distance between the source and observer to be $R_c=100\unit{Mpc}$, and set the impact parameter of the LOS at $b=1\unit{pc}$.
These choices yield a time delay between the null signal and the late-time tail radiation (for a point mass) of $\tau\approx 2.29\unit{s}$ so $\Delta t'\ll \tau$ for $\lambda=100\unit{m}$ radiation, as required.

\paragraph{Solar System}

On the scale of the solar system, we adopt parameters based on the Earth-Moon system.
We choose  $M$ to be the mass of the Moon, the radius of the mass distribution to be the lunar radius ($a=1737.1\unit{km}$), the distance between the source and observer to be twice that between the Earth and Moon ($R_c=7.688\sci{5}\unit{km}$), and a time delay between the null signal and the late-time tail radiation (for a point mass) of $\tau=0.1\unit{s}$.
This choice corresponds to an impact parameter of $b=1.0849\sci{5}\unit{km}$.

\subsubsection{Approximations}
\label{subsec:scalar-approximations}

The field calculated from the Green's function~(\ref{eq:field-scalar}) using a point-charge source~(\ref{eq:J-scalar}) reduces to
\begin{equation}
  \field(x) = \int \dderiv t' q(t') G(x,t',\vec x'_c),
\end{equation}
where we have used the delta function to perform the spatial integral.
In the calculations for a point mass reported above we modeled a charge distribution as a single pulse.
The period of a pulse, $\tau_p$, and the length of the source signal, $\tau_s$, can be decoupled.
We can instead think of the source as being made up of multiple point charges all located at $\vec x'_c$ and each pulsing at different times with the same period, $\tau_p$.
Let the time of a each charge pulse be centered at some time $t_0' \in [t'_{01},t'_{02}]$.
This means that each individual charge only contributes to the signal in the time interval $t'\in[t_0'-\Delta t', t_0'+\Delta t']$.
Further, let the source have a normalized amplitude profile, $s(t_0')$, so that
\begin{equation}
  \int_{t'_{01}}^{t'_{02}} \dderiv t'_0 s(t'_0) = 1.
\end{equation}
The field is then calculated from the convolution
\begin{equation}
  \field(x) = \int \dderiv t'_0 s(t'_0) \int_{t'_0-\Delta t'}^{t'_0+\Delta t'} \dderiv t' q(t'-t_0') G(x,t',\vec x'_c).
\end{equation}
When the period of the pulse is very short compared to the middle time, this expression can be separated by Taylor expanding to leading order in $t'$,
\begin{equation}
  \field(x) \approx \frac{1}{m!} \left( \int_{-\Delta t'}^{+\Delta t'} \dderiv t' q(t') t'^m \right) \left( \int_{t'_{01}}^{t'_{02}} \dderiv t' s(t'_0) \frac{\partial^m G}{\partial t'^m}(x,t'_0,\vec x'_c) \right).
  \label{eq:field-scalar-Taylor}
\end{equation}
Here ``leading order'' means the smallest non-negative integer, $m$, such that the first integral in~(\ref{eq:field-scalar-Taylor}) is non-zero.

In this work we will only be considering systems where this approximation holds.
Further, we will restrict ourselves to a simple top-hat profile,
\begin{equation}
  s_{\textrm{top-hat}}(t'_0) = \frac{1}{t'_{02}-t'_{01}}.
\end{equation}

\subsubsection{Results}
\label{subsec:scalar-results}

\begin{figure}
    \centering
    \begin{subfigure}[t]{0.475\textwidth}
        \centering
        \includegraphics[width=\textwidth]{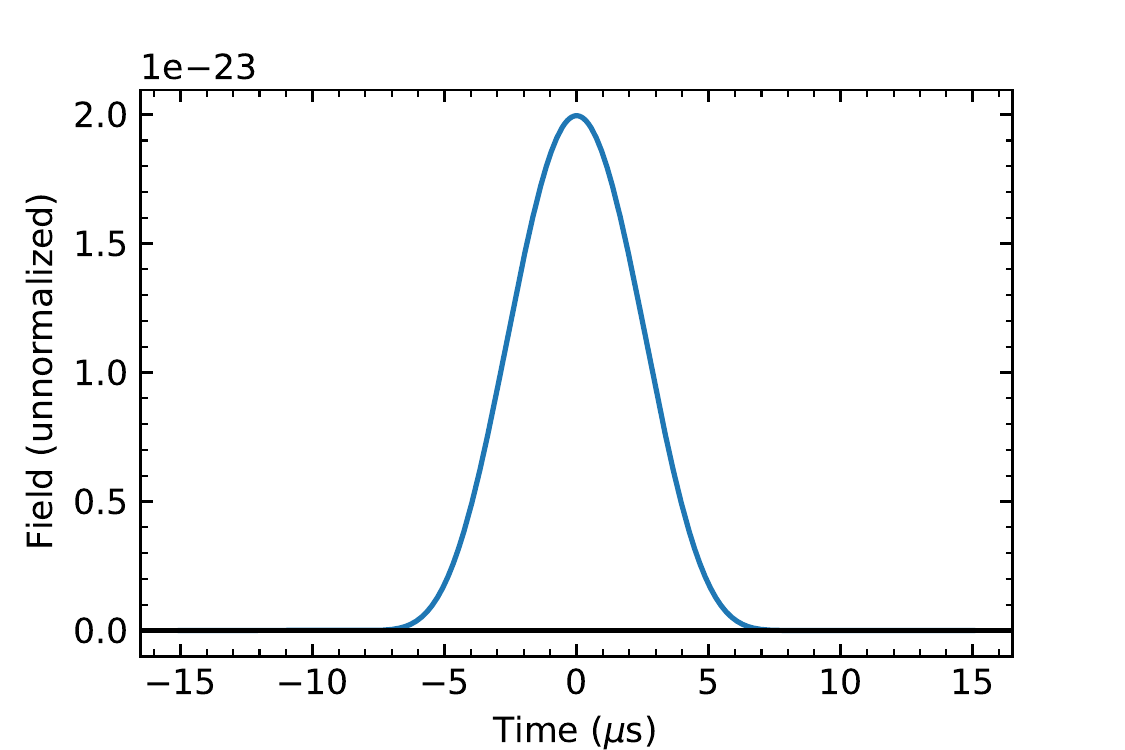}
        \caption{Field as a function of time.}
        \label{fig:CosmoStarSimpleField}
    \end{subfigure}
    \hfill
    \begin{subfigure}[t]{0.475\textwidth}
        \centering
        \includegraphics[width=\textwidth]{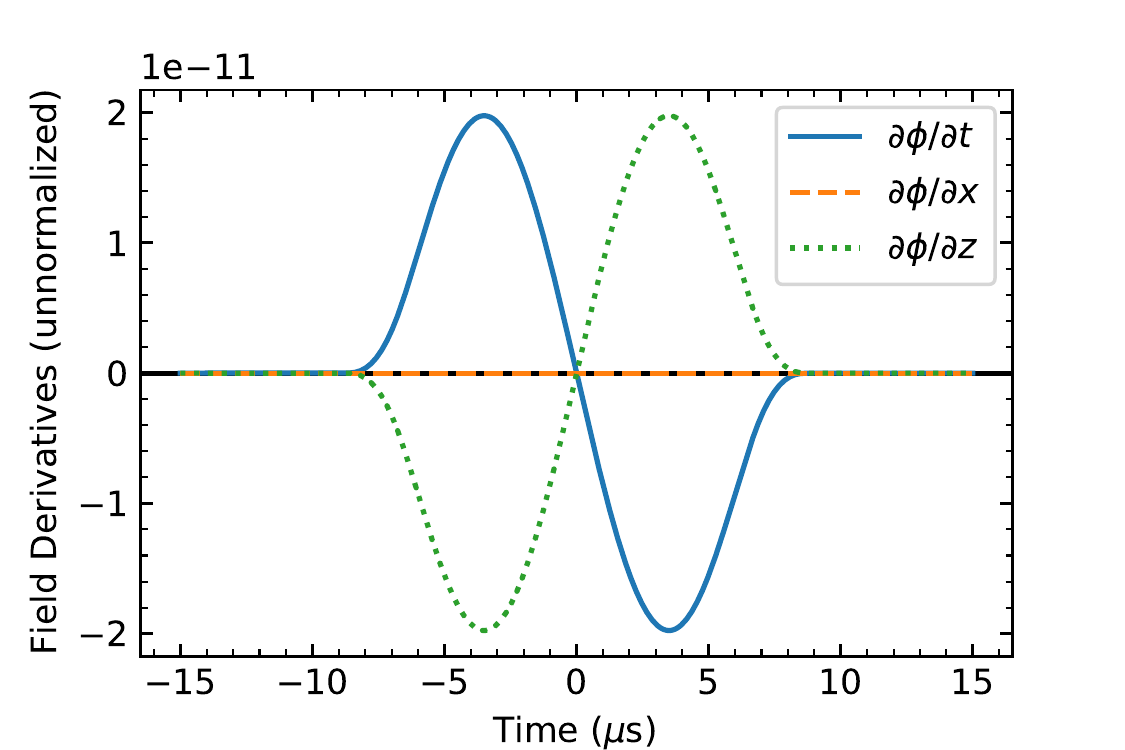}
        \caption{Field derivatives as a function of time.
        (By symmetry $\partial\field/\partial y=0$ and is not shown.)}
        \label{fig:CosmoStarSimpleDerivs}
    \end{subfigure}
    \vskip\baselineskip
    \begin{subfigure}[t]{0.475\textwidth}
        \centering
        \includegraphics[width=\textwidth]{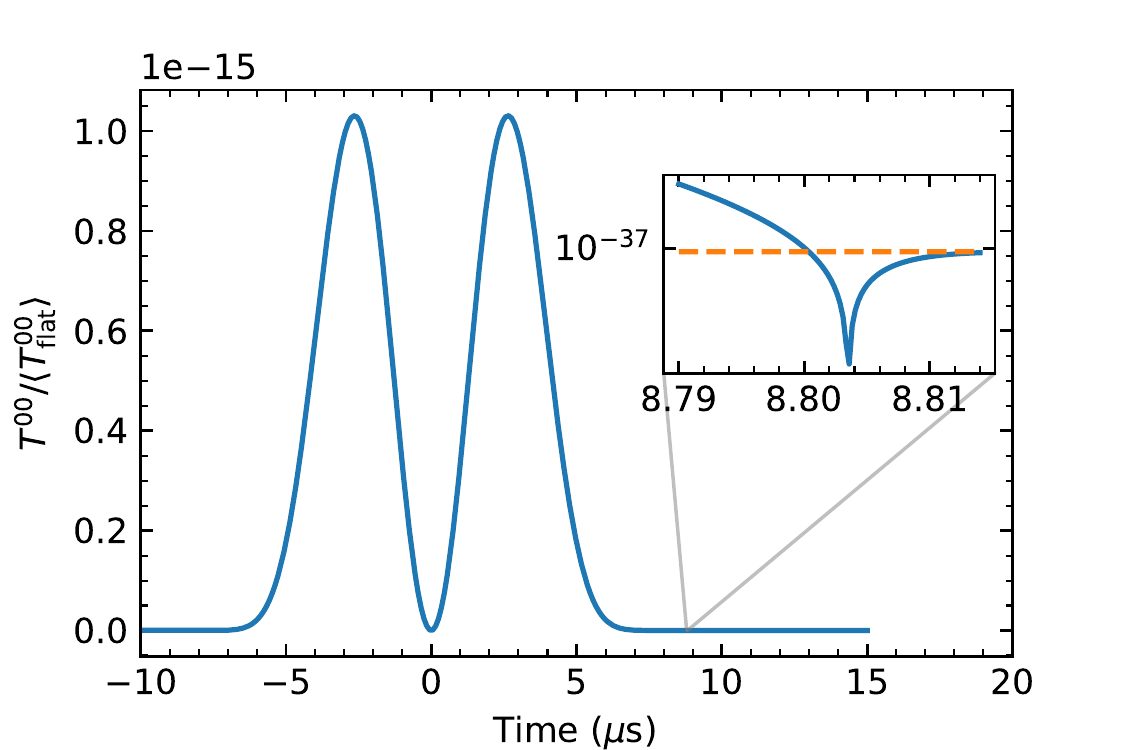}
        \caption{Energy ratio at the observer as a function\\ of time.}
            \label{fig:CosmoStarSimpleEnergy}
    \end{subfigure}
    \begin{subfigure}[t]{0.475\textwidth}
        \centering
        \includegraphics[width=\textwidth]{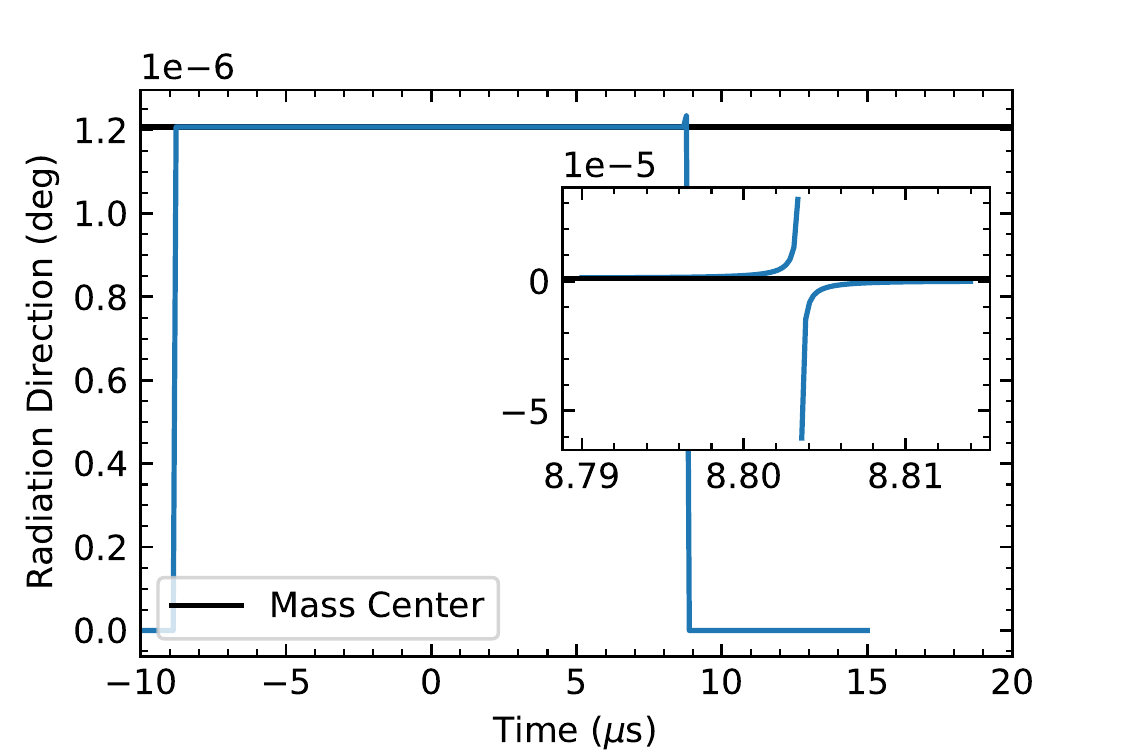}
        \caption{Direction of the incoming radiation as seen by the observer.}
        \label{fig:CosmoStarSimpleDirn}
    \end{subfigure}
    \caption{The scalar tail of the simple pulse for the cosmological-star inspired parameters.
    The insets in the bottom two panels zoom in on the transition from the middle-time tail to the late-time tail.
    The dashed line in the bottom left panel inset shows the value that would be found for the energy ratio from the late-time tail of a point-mass perturber, confirming that energy in the late-time tail is non-zero.
    From the bottom right panel we see that the during the middle-time the radiation appears to come from the perturber, whereas at the late-time it appears to come from the source.
    The $x$-axis shows the time shifted so that $t=0$ is the time when the middle of the pulse emitted at the source would propagate at the speed of light to the center of the perturber and then arrive at the observer.
    }
    \label{fig:CosmoStarSimple}
\end{figure}

 \begin{figure}
    \centering
    \begin{subfigure}[t]{0.475\textwidth}
        \centering
        \includegraphics[width=\textwidth]{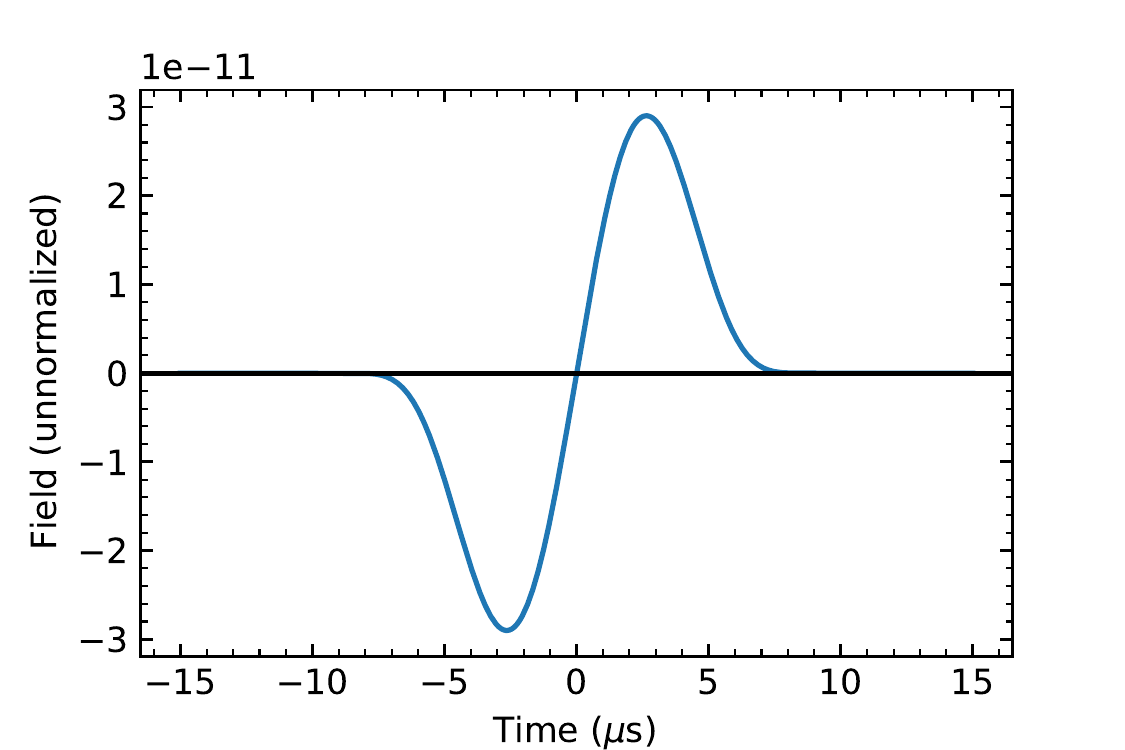}
        \caption{Field as a function of time.}
        \label{fig:CosmoStarSineField}
    \end{subfigure}
    \hfill
    \begin{subfigure}[t]{0.475\textwidth}
        \centering
        \includegraphics[width=\textwidth]{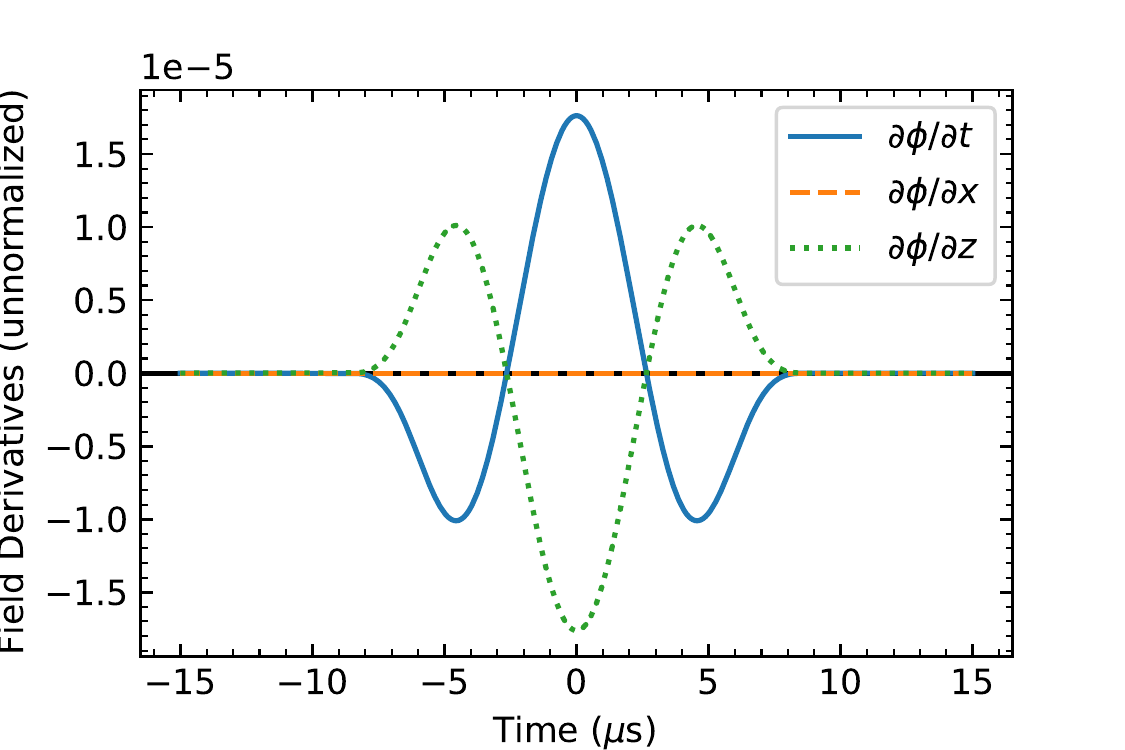}
        \caption{Field derivatives as a function of time.}
        \label{fig:CosmoStarSineDerivs}
    \end{subfigure}
    \vskip\baselineskip
    \begin{subfigure}[t]{0.475\textwidth}
        \centering
        \includegraphics[width=\textwidth]{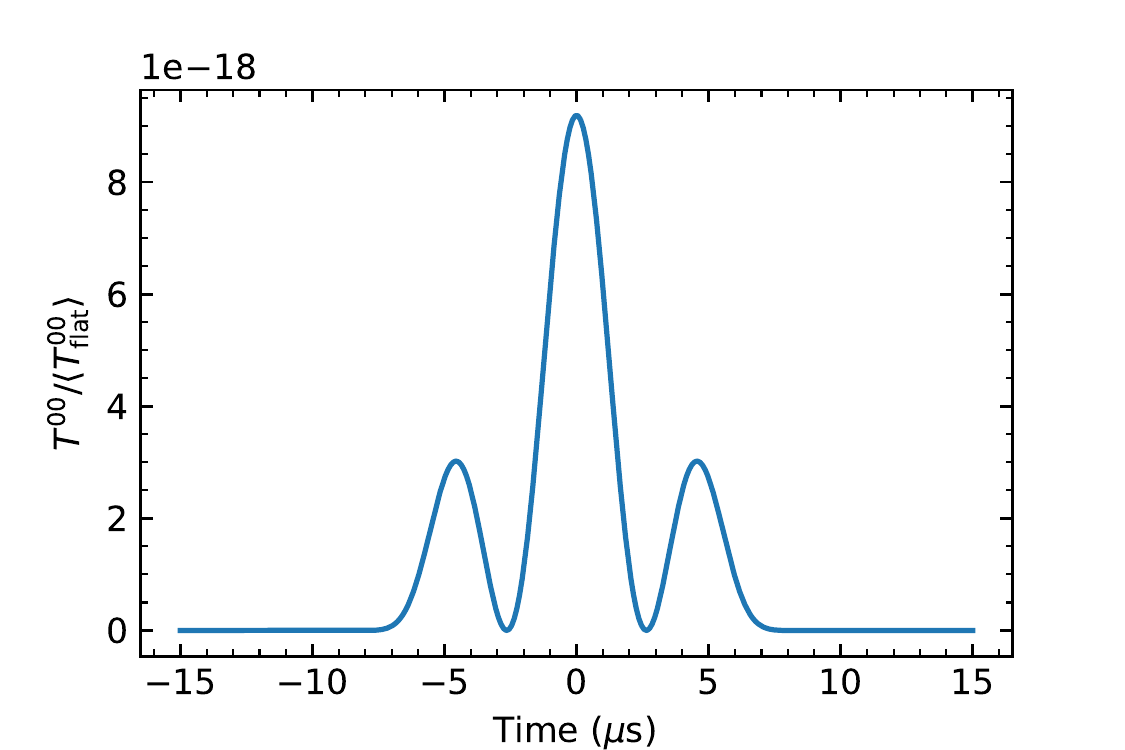}
        \caption{Energy ratio at the observer as a function\\ of time.}
        \label{fig:CosmoStarSineEnergy}
    \end{subfigure}
    \begin{subfigure}[t]{0.475\textwidth}
        \centering
        \includegraphics[width=\textwidth]{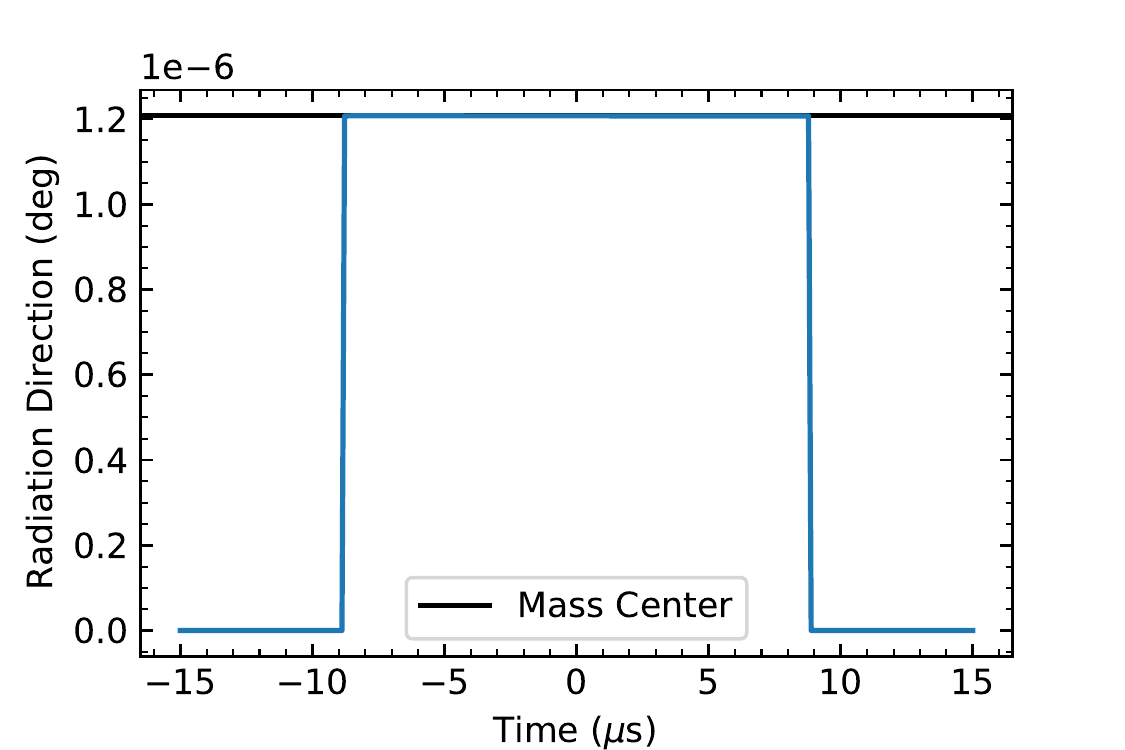}
        \caption{Direction of the incoming radiation as seen by the observer.}
        \label{fig:CosmoStarSineDirn}
    \end{subfigure}
    \caption{The scalar tail of the sine pulse for the cosmological-star inspired parameters.
    See figure~\ref{fig:CosmoStarSimple} for details.}
    \label{fig:CosmoStarSine}
\end{figure}

 \begin{figure}
    \centering
    \begin{subfigure}[t]{0.475\textwidth}
        \centering
        \includegraphics[width=\textwidth]{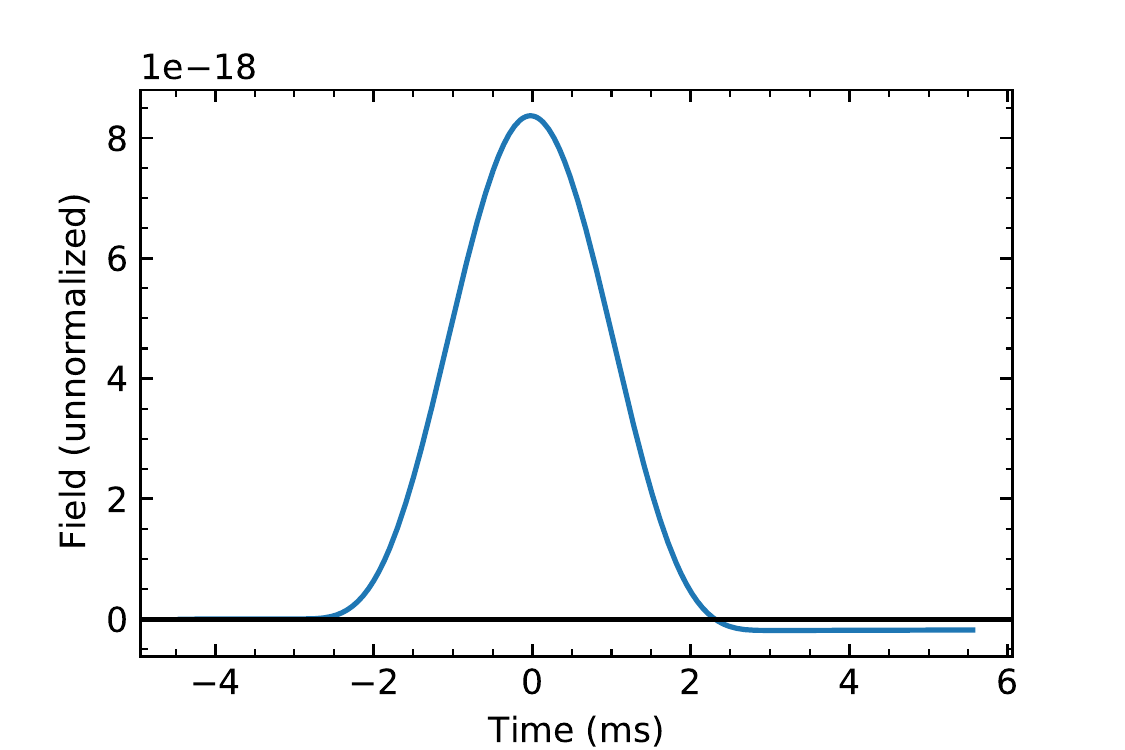}
        \caption{Field as a function of time.}
        \label{fig:LunarSimpleField}
    \end{subfigure}
    \hfill
    \begin{subfigure}[t]{0.475\textwidth}
        \centering
        \includegraphics[width=\textwidth]{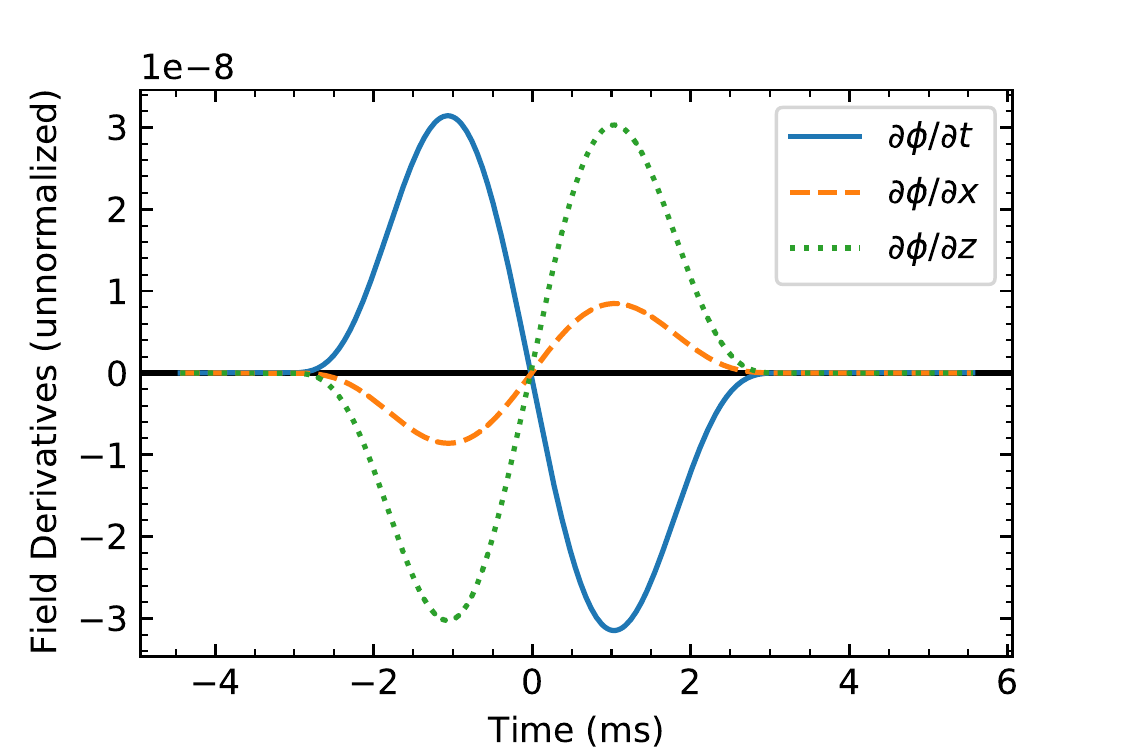}
        \caption{Field derivatives as a function of time.}
        \label{fig:LunarSimpleDerivs}
    \end{subfigure}
    \vskip\baselineskip
    \begin{subfigure}[t]{0.475\textwidth}
        \centering
        \includegraphics[width=\textwidth]{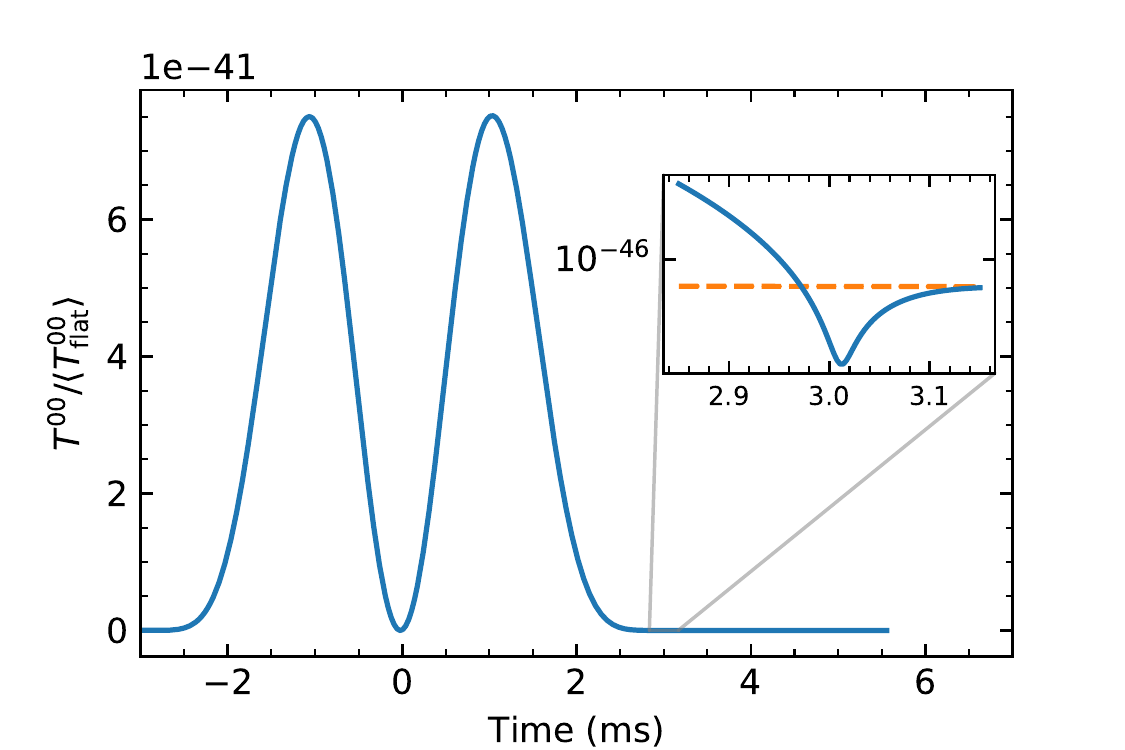}
        \caption{Energy ratio at the observer as a function\\ of time.}
        \label{fig:LunarSimpleEnergy}
    \end{subfigure}
    \begin{subfigure}[t]{0.475\textwidth}
        \centering
        \includegraphics[width=\textwidth]{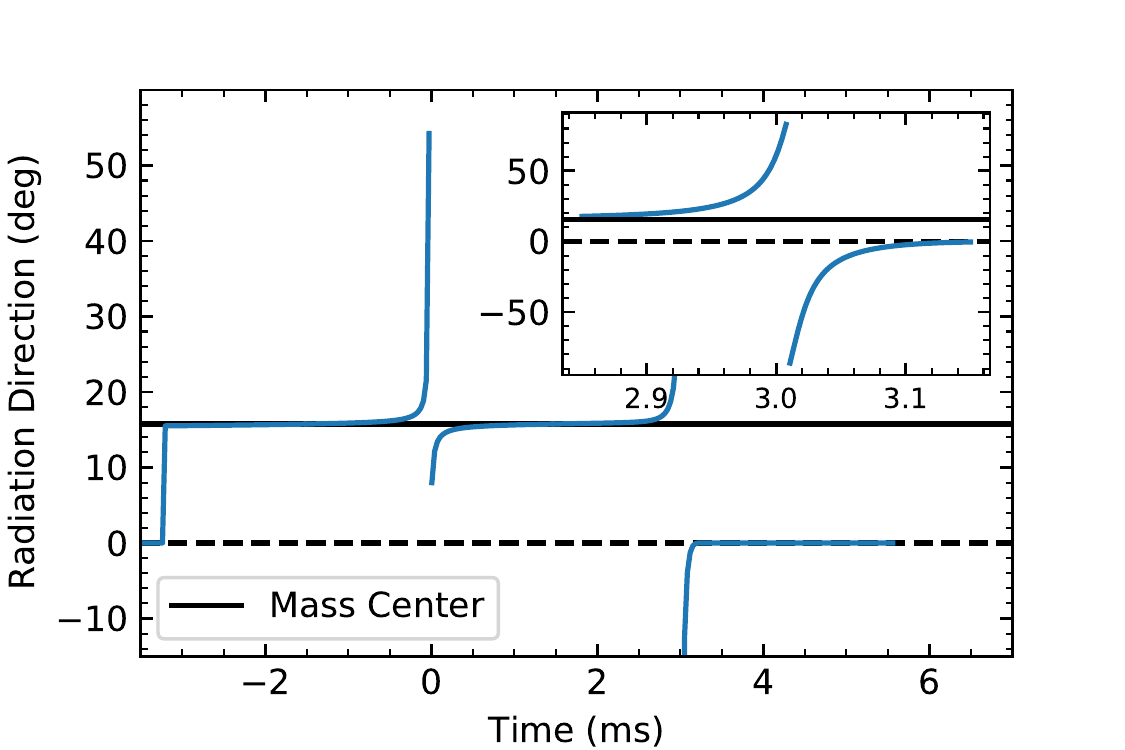}
        \caption{Direction of incoming radiation as seen by the observer.}
        \label{fig:LunarSimpleDirn}
    \end{subfigure}
    \caption{The scalar tail of the simple pulse for the lunar inspired parameters.
    See figure~\ref{fig:CosmoStarSimple} for details.}
    \label{fig:LunarSimple}
\end{figure}

 \begin{figure}
    \centering
    \begin{subfigure}[t]{0.475\textwidth}
        \centering
        \includegraphics[width=\textwidth]{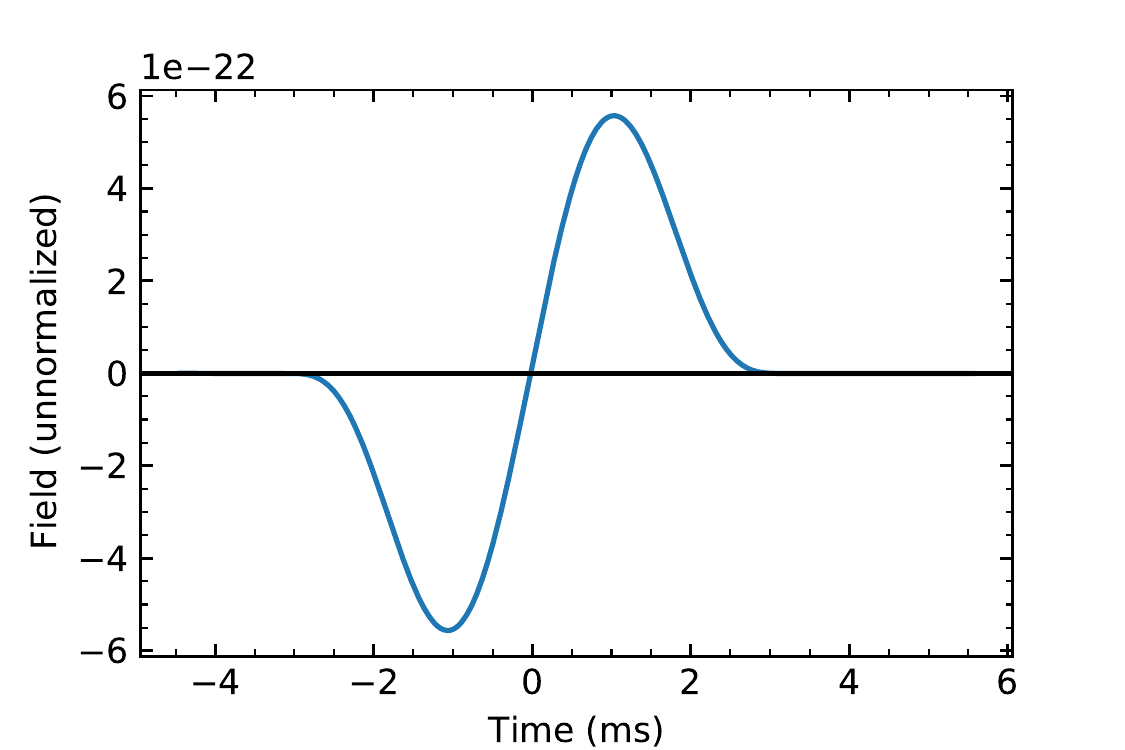}
        \caption{Field as a function of time.}
        \label{fig:LunarSineeField}
    \end{subfigure}
    \hfill
    \begin{subfigure}[t]{0.475\textwidth}
        \centering
        \includegraphics[width=\textwidth]{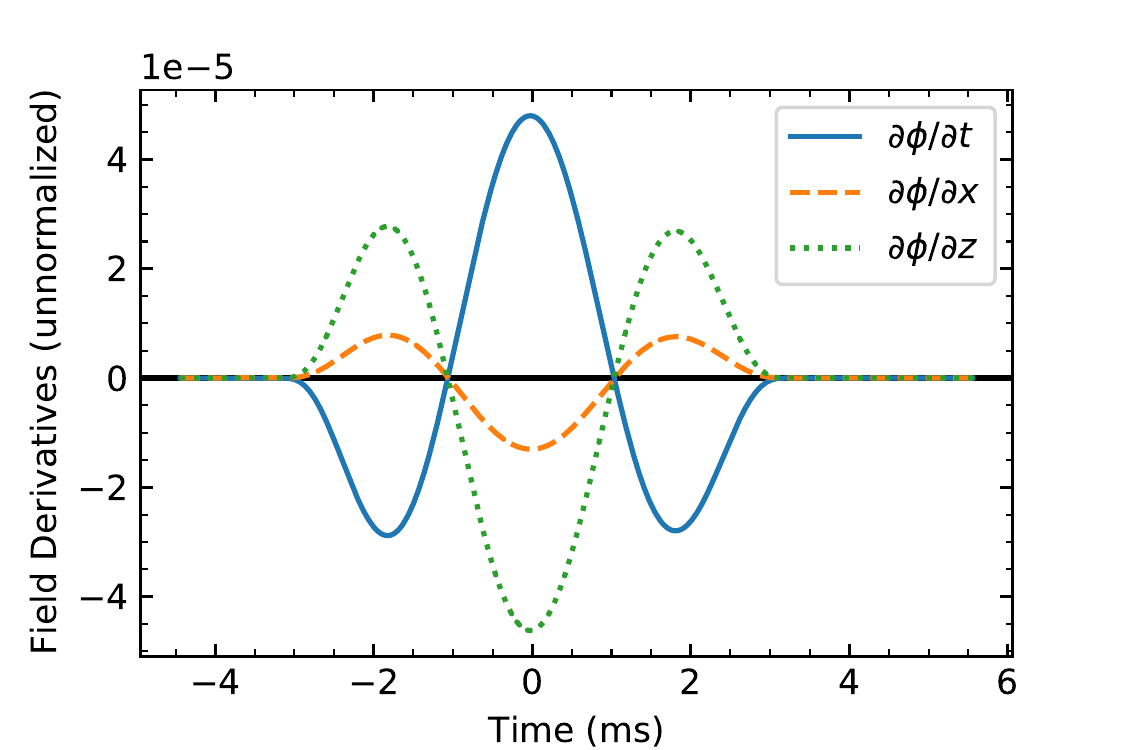}
        \caption{Field derivatives as a function of time.}
        \label{fig:LunarSineeDerivs}
    \end{subfigure}
    \vskip\baselineskip
    \begin{subfigure}[t]{0.475\textwidth}
        \centering
        \includegraphics[width=\textwidth]{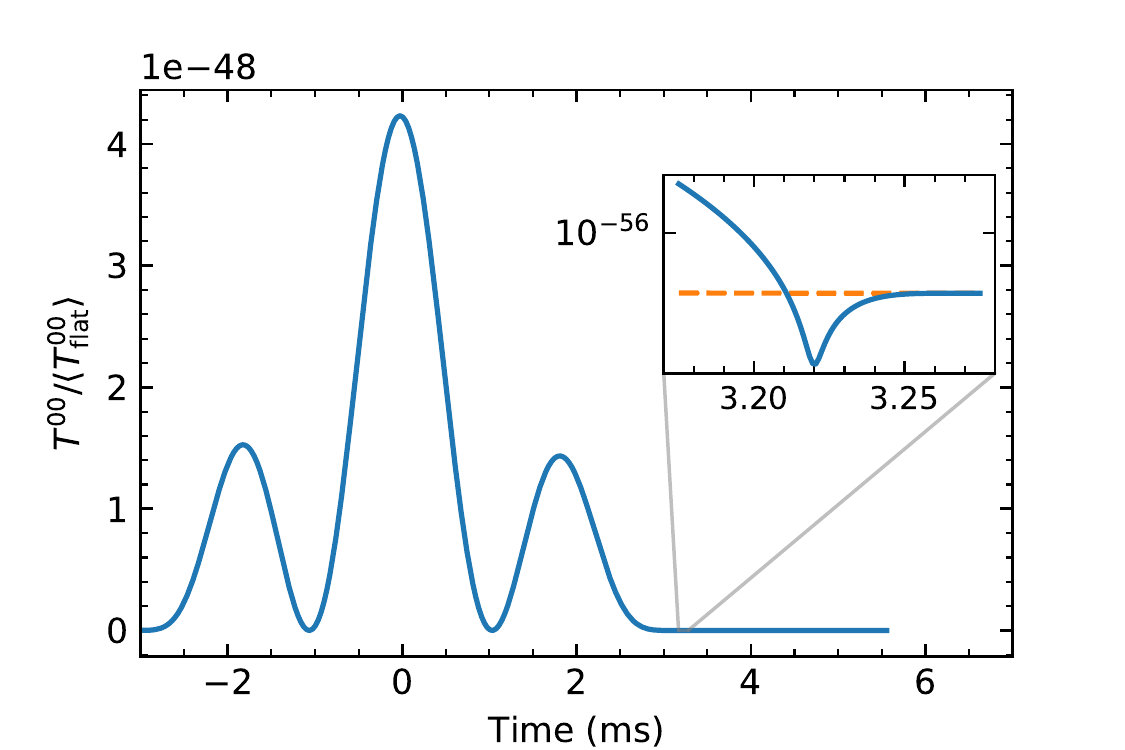}
        \caption{Energy ratio at the observer as a function\\ of time.}
        \label{fig:LunarSineEnergy}
    \end{subfigure}
    \begin{subfigure}[t]{0.475\textwidth}
        \centering
        \includegraphics[width=\textwidth]{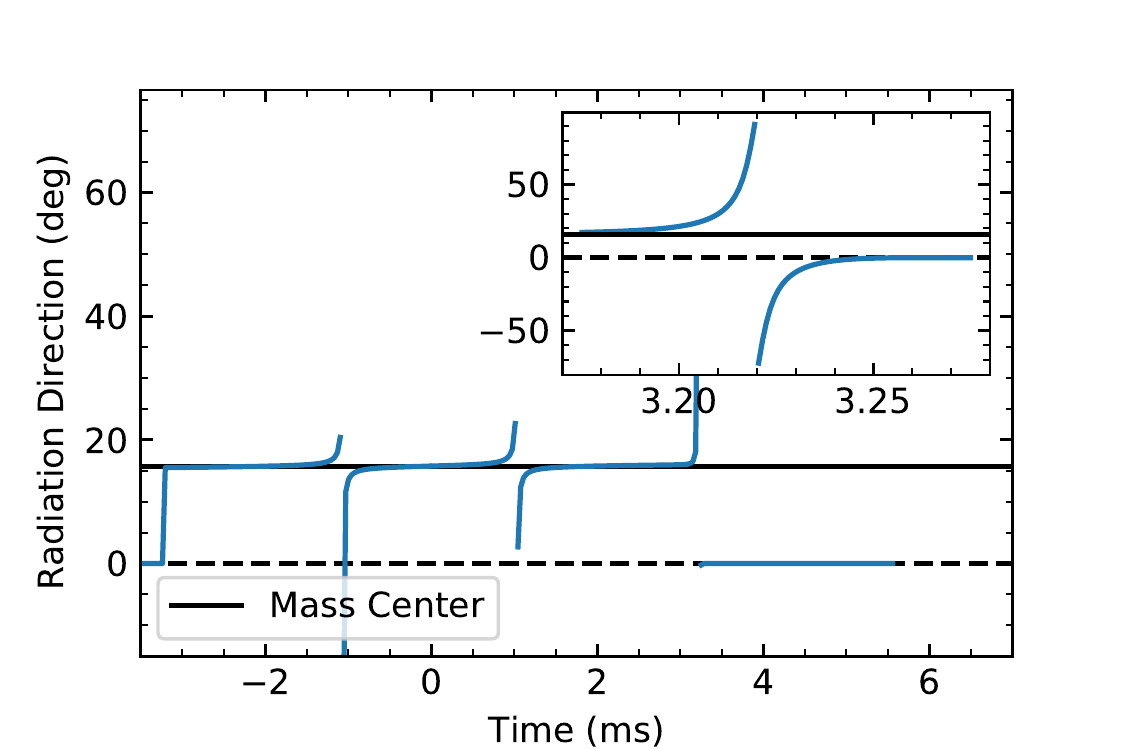}
        \caption{Direction of incoming radiation as seen by the observer.}
        \label{fig:LunarSineDirn}
    \end{subfigure}
    \caption{The scalar tail of the sine pulse for the lunar inspired parameters.
    See figure~\ref{fig:CosmoStarSimple} for details.}
    \label{fig:LunarSine}
\end{figure}

As has already been noted, the work described in this paper does not try to explore the full range of possible geometries (for the observer, perturber, and source), the properties of the perturber (its mass and internal structure), nor the properties of the source (charge densities and spectra).
Rather, it seeks to identify certain specific phenomena through analytic and approximate analytic results, and through a small selection of examples described above.

In figures \ref{fig:CosmoStarSimple} and \ref{fig:CosmoStarSine},
we present the full signal (early-, middle-, and late-time tails) for the cosmological-star parameters probed by a simple pulse and by a sine pulse.
The early-time piece vanishes, at least to leading order, while the late-time piece does not  ---  just as for a point perturber it goes to zero as an inverse power of $t$,  but does not vanish exactly.
Most notable, and previously unappreciated, is the middle-time tail.

The field and its derivatives (the top two panels in figures \ref{fig:CosmoStarSimple} and \ref{fig:CosmoStarSine}) show the importance of the middle piece compared to the late-time tail.
Interestingly, when calculating the energy density, the answer seems to be very sensitive to the smoothness of the density profile as well as to the form of the source pulse.
For example, for the simple pulse (figure \ref{fig:CosmoStarSimpleEnergy}) we see two peaks, while for the sine pulse (figure \ref{fig:CosmoStarSineEnergy}) we see three peaks in the observable energy.
The number of peaks observed is related to a combination of the shape of the mass distribution and the form of the source signal.
While we defer a detailed exploration to future work, this suggests the potential of the tail to probe the mass distribution in a way that would allow it to be reconstructed, even though the LOS is well exterior to the perturbation.

After the signal has finished probing the mass distribution, as it transitions from middle to late times, the tail displays a tiny ``glitch'' --- a valley as shown in figure \ref{fig:CosmoStarSimpleEnergy}, for which we have no physical explanation.
No similar glitch develops in the $T^{00}$ at the transition from the early time to the middle time. 
We cannot guarantee that the existence or not of these glitches is robust against  higher order corrections in small quantities.

As for the point perturber, the late-time tail (unlike the early-time) is non-zero, but much smaller than the middle-time tail.
This also means that for middle times the interior piece in the $A$ function dominates the exterior piece (i.e.\ the extension of the late-time tail into the middle time), making the contribution from the exterior piece comparably negligible.

In figures \ref{fig:CosmoStarSimpleDirn} and \ref{fig:CosmoStarSineDirn} we plot the direction from which the observer infers the tail radiation is arriving (calculated as $\theta=\arctan(T^{0\perp}/T^{0z})$, where $T^{0\perp}=T^{0x}$  due to the choice of placing the perturber on the $x$-axis, which guarantees that $T^{0y}=0$).
In these radiation-direction plots, we discover that for the middle times the radiation appears to be coming from the mass distribution, consonant with the interpretation of the tail as being due to ``scattering'' from the perturber.
However the late-time tail appears to be coming directly from the source,\footnote{
Note that at leading order we cannot resolve the perturber so the ``mass center'' includes the entire mass distribution.}
just as it does in the null signal.
This does not appear to have been remarked on in past works.
Mathematically, it is unsurprising as the late-time Green's function is independent of the location of the perturber;
however, it would seem to undermine the interpretation of the late-time part of the tail as scattering from the geometry --- i.e.\ from the perturbation to the geometry sourced by the perturber.
To us, the mystery remains as to the physical origin of the late-time tail.

The results for the ``lunar'' parameters perturber presented in figure \ref{fig:LunarSimple} for the simple pulse and figure \ref{fig:LunarSine} for the sine pulse are qualitatively similar to those of the cosmological-star parameters.
We have, however, taken the opportunity in panel (d) of figures~\ref{fig:LunarSimple} and \ref{fig:LunarSine} to zoom in on the period when the apparent source of the radiation changes from the perturber to the source.  
This change happens quite abruptly near the end of the middle time.  The direction of origin of the signal appears not to the take the short angular path from the perturber to the source, but rather to ``whip around'' the observer in almost a full circle --- appearing in the figures like a divergence in the radiation direction. 
We note that in the case of the sine pulse (figure~\ref{fig:LunarSine}) there are multiple moments when this whip-around occurs, though here the signal starts and ends at the perturber.
We caution the reader that this behavior always coincides with a moment when $T^{0x}$ and $T^{0z}$ both pass through zero.
No radiation actually arrives from directly behind the observer.
Moreover, it means that this result is not necessarily robust against higher order corrections in small quantities.
What \emph{is} robust is that in the  middle time the observer detects the signal coming from the perturber, and in the late time it comes from the source.

The principle difference between the cosmological-star and lunar parameters is the overall magnitude of the middle-time tail, which is much larger for the cosmological-star parameters.
The amount of radiation in the middle-time tail depends on all of the parameters of the system --- many of them in ways we have barely begun to disentangle.
In the lower-left panels of figures \ref{fig:CosmoStarSimple}--\ref{fig:LunarSine} we have compared the energy density in the  middle-time tail to the average energy density in the null radiation.
While that ratio is always small, for the cosmological-star parameters it is not so terribly small --- reaching $10^{-15}$ for the simple pulse.
A more exhaustive search may reveal conditions under which the ratio is even larger.

If there were massless scalar radiation, there is the potential for a detectable signal.
While the Standard Model of particle physics does not have any massless (or nearly massless) scalars, they are common components of Beyond the Standard Model (BSM) physics, especially axions and axion-like particles (ALPs).
Moreover, most or all weak-field perturbers would be nearly fully transparent to axions or ALPs --- unlike photons.
If these particles mixed with photons (and axions do), one could imagine visible electromagnetic tails resulting.

\begin{figure}
  \centerline{\includegraphics{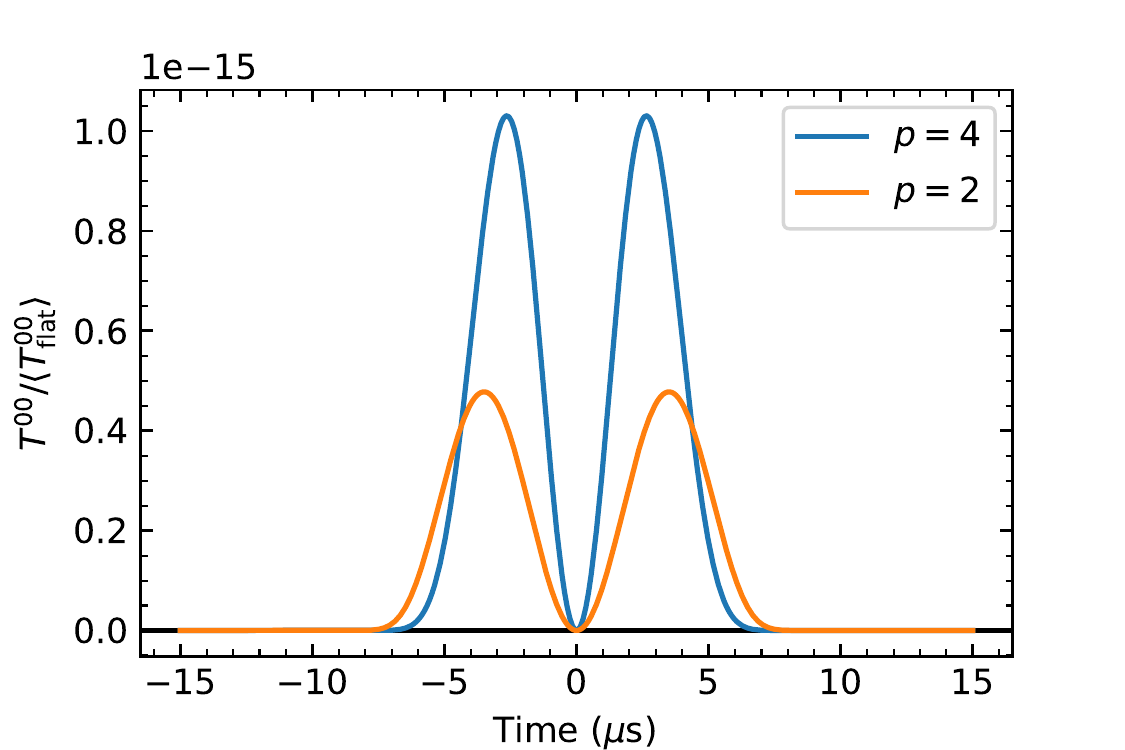}}
  \caption{Mass distribution density dependence.
  The effect on the energy density of changing the power in the density profile of the mass distribution~(\ref{eq:mass_distn}) for the cosmological-star inspired parameters with the simple pulse.
  Increasing $p$ makes the gradient of the mass distribution larger which corresponds to a larger energy density during the middle-time tail.
  The time on the $x$-axis is shifted as in figure~\ref{fig:CosmoStarSimple}.
  See the text for more details.
  }
  \label{fig:CosmoStarSimple-rho-dependence}
\end{figure}

\begin{figure}
  \centerline{\includegraphics{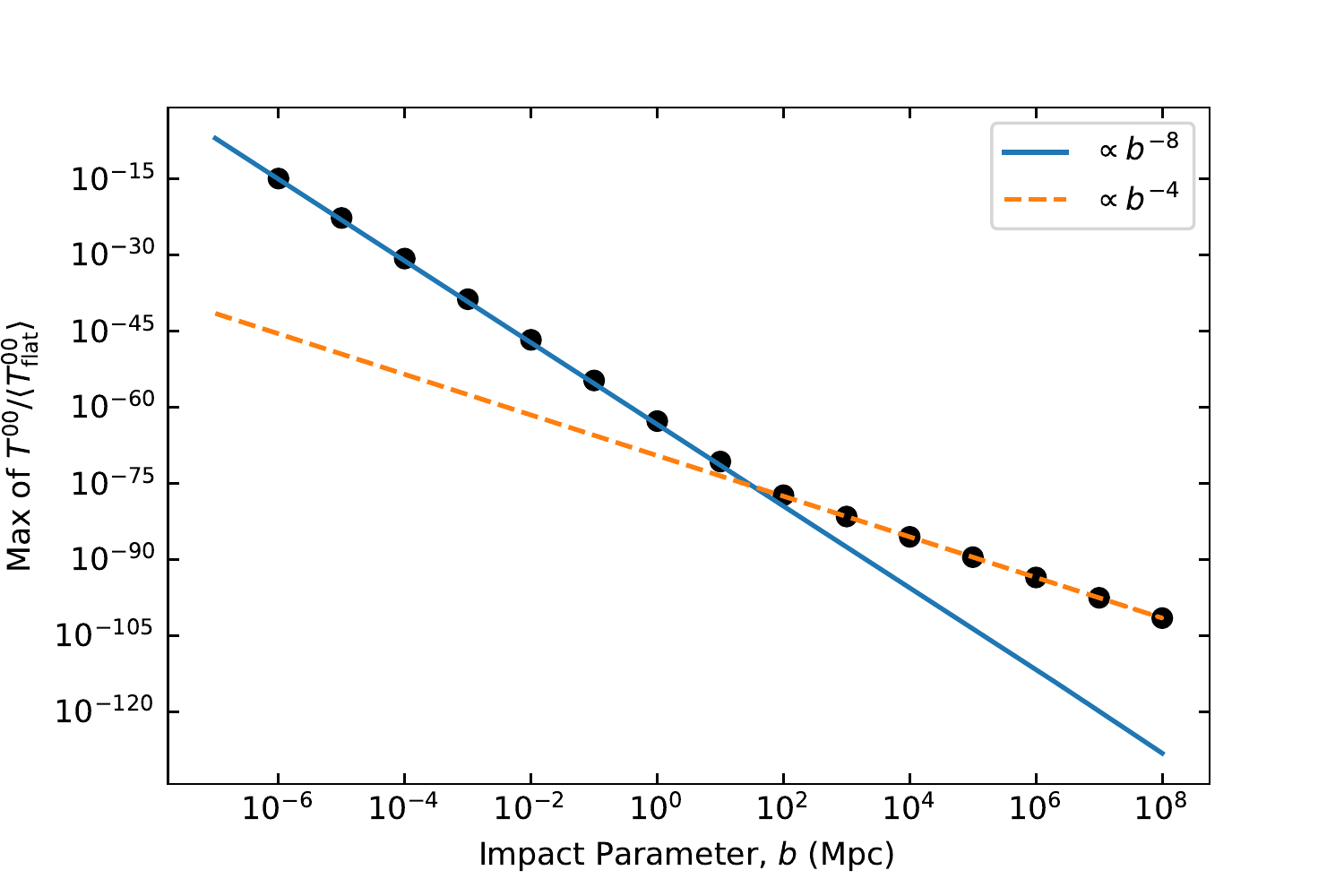}}
  \caption{Scaling of the energy observed in the middle-time tail as a function of the impact parameter.
  The ratio of the peak energy density in the middle-time tail to the average of the energy density in flat spacetime versus the impact parameter is shown for the cosmological-star inspired parameters with the simple pulse.
  Two different scaling behaviors are seen.
  See the text for more details.
  }
  \label{fig:CosmoStarSimple-b-scaling}
\end{figure}

Although an exhaustive search of parameter space has not been performed and is reserved for future work, a preliminary exploration provides some insights into the expected behaviors of the middle and late-time contributions to the tails.
We explore below the dependence of the tails on the pulse frequency (through the period, $\tau_p=2\Delta t'$), the perturbing mass distribution ($p$, $a$, and $M$), and the system geometry ($b$).

\paragraph{Dependence on pulse frequency}

Frequency dependence of our results enters through the form of the time-varying charge, $q(t')$.
As shown above for the approximations used in this work~(\ref{eq:field-scalar-Taylor}), the calculation of the field factors into a term that depends only on the Green's function and a term that depends only on the charge distribution.
This allows for an analytic calculation of the frequency dependence.
For two different frequencies the ratio of the energy densities is determined from
\begin{equation}
    \frac{T^{00}_1}{T^{00}_2} = \left. \left( \int_{-\Delta t'_1}^{+\Delta t'_1} \dderiv t'\, q(t') t'^m \right)^2 \middle/ \left( \int_{-\Delta t'_2}^{+\Delta t'_2} \dderiv t'\, q(t') t'^m \right)^2 \right. .
\end{equation}
The required integrals can be calculated analytically for the charge distributions considered in this work.
We find
\begin{align}
    \frac{T^{00,\mathrm{simple}}_1}{T^{00,\mathrm{simple}}_2} &= \left( \frac{\Delta t'_1}{\Delta t'_2} \right)^2, \nonumber \\
    \frac{T^{00}_{n_1,1}}{T^{00}_{n_2,2}} &= \left( \frac{\Delta t'_1}{\Delta t'_2} \right)^4 \left( \frac{n_2}{n_1} \right)^2.
\end{align}
This shows the expected general behavior: as $\Delta t'$ decreases (frequency increases) for the signal pulse, the intensity in the tail decreases.
The fact that the tail is stronger for lower frequencies (higher $\Delta t'$, lower $n$) is previously  known for the late-time tail, but previously unstudied for the middle-time tail.

\paragraph{Dependence on perturbing mass distribution ($M$, $p$, $a$)}

The dependence of the tail on $M$ is simple to leading order --- the field in the tail is proportional to $M$, and therefore the energy is proportional to $M^2$.
Other dependencies are less trivial.

The results in this work have used a density profile~(\ref{eq:mass_distn}) with $p=4$.
This was chosen to ensure that all quantities calculated (in particular the stress-energy tensor) were smooth at the edges of the mass distribution for both of the pulses studied here.
(For the simple pulse, $p=2$ is sufficient to ensure smoothness.)
The density distribution affects only the calculation of $A_{\text{interior-x}}$, changing the order of polynomial that appears in~(\ref{eq:Ainterior-x}).
The dependence of the energy density on $p$ is shown in figure~\ref{fig:CosmoStarSimple-rho-dependence} for the cosmological-star parameters with the simple pulse.
As $p$ increases, while the size, $a$, and total mass, $M$, are held fixed, the density distribution becomes more peaked, leading
to larger density gradients.
This results in an increase in both the peak energy in the  middle-time tail and the total energy in the middle-time tail.
The dependence is approximately linear: going from $p=2$ to $p=4$ increases the peak energy by slightly more than a factor of two.
We have not made a sufficient exploration of this phenomenon to generalize broadly beyond these very particular pair of distributions.

Finally, probing the dependence of the field or the stress-energy tensor on the radius of the perturber, $a$, is not straightforward, since the result is sensitive to the value of other parameters.
On the one hand, our approximations require that $a\ll b$, so studying the behavior as $a$ increases also requires increasing $b$.
On the other hand, our approximations require that $a\gg M$ to ensure that $|\Phi|\ll 1$, so studying the behavior as $a$ decreases requires decreasing $M$.
Study of the scaling with $a$ will be left to future work where a more complete exploration of parameter space is performed.

\paragraph{Dependence on system geometry ($b$)}

We have  made a preliminary exploration of the peak energy density in the middle-time tail as a function of the impact parameter $b$.
Figure~\ref{fig:CosmoStarSimple-b-scaling} shows this scaling for the cosmological-star parameters with the simple pulse where only $b$ is allowed to vary.
The scaling roughly follows two power-laws: for small impact parameters there is a $b^{-8}$ dependence, whereas for large impact parameters there is a $b^{-4}$ dependence.
The location of the break between ``small'' and ``large'' $b$ depends on other parameters.
The steep dependence at small impact parameters is enticing, suggesting that the middle-time tail contribution could be much larger than we have shown.
We cannot push $b$ to smaller values without breaking some of our simplifying assumptions, necessitating new approaches to calculating the behavior.
An exploration of even smaller impact parameters has hinted at a flattening of the curve precisely in the regime where our approximations break down, but therefore cannot be trusted.
Determining the behavior at very small impact parameters would be exciting and is reserved for future work.

We have not made any attempt to relax other simplifying assumptions --- such as moving the perturber so that it is not positioned equidistant from the source and the observer.

\paragraph{Summary of scalar results}

\emph{The energy carried by the middle-time tail is much larger than the energy in the late-time tail --- the tail-to-null ratio computed for the point-mass perturber is a significant underestimate of both the total tail radiation and its peak intensity.}
While this may have been unanticipated, there was no good reason to trust the point-perturber calculation, since close to the delta-function one necessarily violates the weak-field limit.
The middle-time energy appears to the observer to come from the direction of the perturber, in line with the usual physical interpretation of the tail as being radiation scattered off the perturbed spacetime.
This is an important new result of this paper with potential observational consequences  that carries over to electromagnetic radiation (as we show below), and therefore likely also to gravitational radiation.

The energy in the middle-time tail reveals a structure of peaks and dips that we did not anticipate, nor do we yet understand physically.
The tail signal depends on both the time dependence of the source and the radial dependence of the mass distribution of the perturber.
The factorization of the integrals in~(\ref{eq:field-scalar-Taylor}) is instructive and a calculation aid.
It means, for example, that the  pulse shape (and so, for example, $n$ and $\Delta t'$) has two effects --- it determines the integral over $q(t')$, and it determines the value of $m$, i.e.\ the number of derivatives of the Green's function in the leading-order term.

The middle-time tail lasts until the latest time that a signal can travel at the speed of light from the source to a point within the mass distribution, and then to the observer.
Once the middle-time tail is over, there remains a late-time tail that is identical to the one calculated for the point-mass perturber of the same mass and location.
Although it is characteristically considerably smaller than the middle-time tail, the late-time tail does carry energy and momentum.
The late-time tail appears to the observer to come not from the direction of the perturber but instead from the direction of the source.
It is not clear to us physically why there is a late-time tail when there was no early-time tail, nor why it appears to the observer to come from the source, when the perturber is the proximate cause.

\section{Vector radiation}
\label{sec:vector-radiation}

With the techniques, approximations, and intuition developed in the case of massless-scalar radiation, we can turn our attention to massless-vector radiation: the usual electromagnetic radiation which is of obvious importance.
The exploration of the vector tail radiation closely follows that of the scalar radiation (section~\ref{sec:scalar-radiation}), so the techniques can be applied with little discussion needed.
There are some important differences between the scalar and vector cases: the vector Green's function has multiple components, and includes a contribution from the Ricci tensor~(\ref{eq:G-vector}); the relevant fields are the electric and magnetic fields;
it is more convenient to discuss the energy and momentum densities in terms of the Poynting vector, rather than directly in terms of the stress-energy tensor; and as our source we will consider an oscillating electric dipole (since there is no monopole electromagnetic radiation).

\subsection{Vector radiation: the null-cone piece in Minkowski spacetime}
\label{subsec:vector-nullcone}

The null, flat-space results are standard, textbook calculations.
Here we simply report the results for a point-like dipole, $d \ll \Delta t' \ll R_c$, in a coordinate system with its origin at the center of the dipole.

\paragraph{Simple Pulse}

\begin{align}
    \doflat{\vec E}_{\mathrm{simple}} &\approx  -\frac{3 q _0 d (t-R_c)^2}{\pi \Delta t^{\prime 4}} \left(\frac{\sin\theta}{R_c}\right) \unitvec{\theta}, \\
    \doflat{\vec B}_{\mathrm{simple}} &\approx -\frac{3 q _0 d (t-R_c)^2}{\pi \Delta t^{\prime 4}} \left(\frac{\sin\theta}{R_c}\right) \unitvec{\varphi}.
\end{align}

\paragraph{Sine Pulse}

\begin{align}
    \doflat{\vec E}_n &\approx \frac{n^2 \pi q_0 d}{4 \Delta t^{\prime 2}} \left( \frac{\sin\theta}{R_c} \right) \sin\!\left[ \frac{n\pi (t-R_c)}{\Delta t'} \right] \unitvec{\theta}, \\
    \doflat{\vec B}_n &\approx \frac{n^2 \pi q_0 d}{4 \Delta t^{\prime 2}} \left( \frac{\sin\theta}{R_c} \right) \sin\!\left[ \frac{n\pi (t-R_c)}{\Delta t'} \right] \unitvec{\varphi}.
\end{align}

\subsection{Vector tail radiation with point perturber}
\label{subsec:vector-pointperturber}

As in section~\ref{subsec:scalar-pointperturber} for the scalar radiation, we again begin with a point-mass perturber.
Outside the mass distribution the Ricci-tensor contribution is zero every where except at the location of the point mass.
From the vector Green's function~(\ref{eq:G-vector}), it can be shown that
\begin{equation}
    G^{\textrm{point-mass}}_{\mu \nu'} (x,x') = - \frac{\eta_{\mu\nu}}{2\pi} \partial_{t} \partial_{t'} A(x,x').
\end{equation}
Just as in the scalar case, since $\Aearly(x,x')$ is independent of time, so, once again, \emph{there is no early-time tail radiation} at leading order in perturbations,
\begin{equation}
    G^{\mathrm{early}}_{\mu \nu'} (x,x') = 0.
\end{equation}
For the late-time Green's function direct calculation gives
\begin{align}
    \Glate_{\mu\nu'} (x,x') &= - \frac{\eta_{\mu\nu}}{2\pi} \partial_{t} \partial_{t'} \Alate(x,x')
        = - \frac{2M}{\pi} \frac{t-t'}{\left[ (t-t')^2 - |\vec x-\vec x'|^2 \right]^2} \eta_{\mu\nu}, \\
        &= \eta_{\mu\nu} \Glate(x,x'), \nonumber 
\end{align}
where in the last line we exhibit the simple relationship to the late-time scalar Green's function~(\ref{eq:Glate-scalar}).

The structure of the late-time Green's function allows us to write the electric and magnetic fields~(\ref{eq:field-vector}) in terms of the components of the Maxwell field-strength tensor in a simple manner:
\begin{align}
   \dolate{F}_{0i} &= - \int \dderiv^4 x' \left( J^{i'} \partial_t + J^{0'} \partial_i \right) \Glate_{00'}(x,x'),
    \label{eq:Efield-late} \\
    \dolate{F}_{ij} &= - \int \dderiv^4 x' \left( J^{j'} \partial_i - J^{i'} \partial_j \right) \Glate_{00'}(x,x').
    \label{eq:Bfield-late}
\end{align}

\subsubsection{Radiation at infinity}

\begin{figure}
  \centerline{\includegraphics{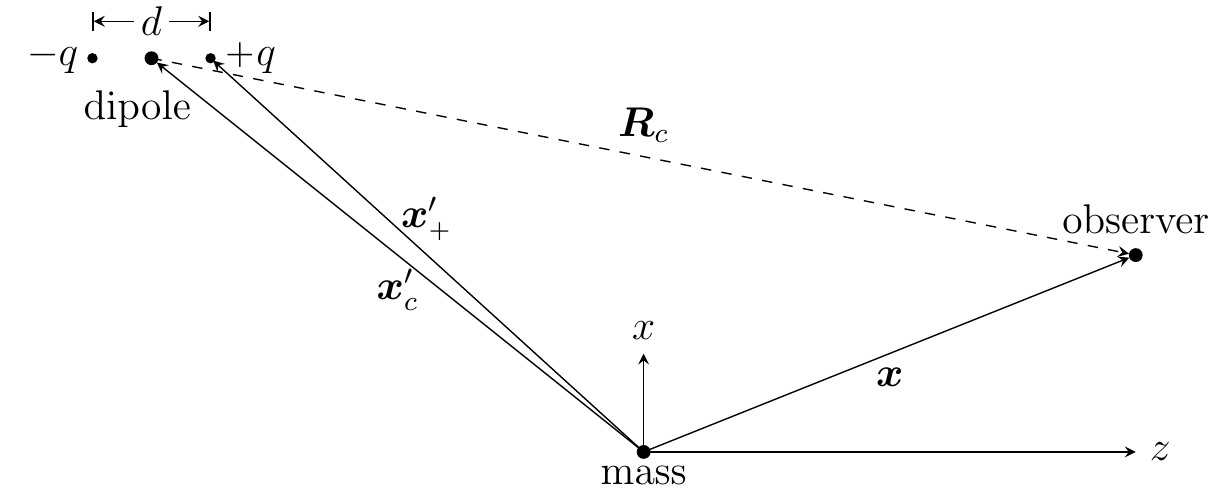}}
  \caption{Coordinate system for a point-mass perturber with a dipole source and an arbitrary observer.
  The coordinate system has its origin at the center of the point mass.
  The center of the dipole and the point mass lie in the $xz$-plane with the dipole aligned along a line parallel to the $z$-axis.
  The observer is located at an arbitrary position.
  Not drawn to scale.
  }
  \label{fig:point_mass-dipole-coordinate-system}
\end{figure}

The study of radiation at infinity is similar to that in section~\ref{sec:scalar-radiation-approximation} for scalar radiation.
The coordinate system we choose is similar, though a dipole has an extra direction associated with it, something that is missing in the monopole scalar source.
Even so, without loss of generality, we can choose the dipole to be aligned along a line parallel to the $z$-axis, and the dipole and perturbing mass distribution to be located in the $xz$-plane, as shown in figure~\ref{fig:point_mass-dipole-coordinate-system}.
The observer is located at an arbitrary location $\vec x=(x,y,z)$ and the center of the dipole is located at $\vec x_c'=(x_c',0,z_c')$.
With this configuration, the components of the current density four-vector are
\begin{align}
    J^0(t',\vec x') &= q(t') \delta(y') \delta(x'-x_c') \left[ \delta\left(z'-z_c'-\frac{d}{2}\right) - \delta\left(z'-z_c'+\frac{d}{2}\right) \right], \\
    J^z(t',\vec x') &= i(t') \delta(y') \delta(x'-x_c') \left[ \Theta\left(z'-z_c'+\frac{d}{2}\right) - \Theta\left(z'-z_c'-\frac{d}{2}\right) \right],
\end{align}
where $i(t')\equiv \dderiv q(t')/\dderiv t$.

\paragraph{Electric and Magnetic Fields}
Plugging the current density into the electric field components~(\ref{eq:Efield-late}) and integrating by parts,
\begin{align}
    \dolate{\vec E} & \approx
     - \frac{48 GM d}{\pi} \left[\unitvec{x}(x-x_c') + \unitvec{y}y \right] (z-z_c') \int \dderiv t' q(t') \frac{T}{(T^2-R_c^2)^4} -  \\
      & \qquad {} - \frac{2 GM d}{\pi} \unitvec{z} \left\{ \left[ q(t') \frac{3T^2+R_c^2}{(T^2-R_c^2)^3} \right|_{t_1'}^{t_2'} - \int \dderiv t' q(t') \frac{8 T[T^2+2R_c^2-3(z-z_c')^2]}{(T^2-R_c^2)^4} \right\}, \nonumber
    \label{eq:Efield-arbitrary-observer}
\end{align}
and similarly the magnetic field calculated from~(\ref{eq:Bfield-late})
\begin{equation}
    \dolate{\vec B} \approx
    \frac{8 GM d}{\pi} [- y \unitvec{x} + (x-x_c') \unitvec{y}] \left\{ \left[ q(t') \frac{T}{(T^2-R_c^2)^3} \right|_{t_1'}^{t_2'} - \int \dderiv t' q(t') \frac{5T^2+R_c^2}{(T^2-R_c^2)^4} \right\},
    \label{eq:Bfield-arbitrary-observer}
\end{equation}
where $T\equiv t-t'$ and $R_c^2\equiv |\vec x-\vec x'|^2 = (x-x'_c)^2 + y^2 + (z-z'_c)^2$.

\paragraph{Simple Pulse}

As an application of the above expressions, consider the simple pulse~(\ref{eq:q-simple}) introduced in the scalar field calculation.
Converting to spherical coordinates with an origin at the center of the dipole, $(R_c,\theta,\varphi)$, and expanding to leading order in $\Delta t'$, we find
\begin{align}
    \dolate{\vec E}_{\mathrm{simple}} &\approx -\frac{128 M q_0 d \Delta t'}{5\pi} \frac{t R_c^2}{(t^2-R_c^2)^4} \sin(2\theta) (\unitvec{x} \cos\varphi + \unitvec{y} \sin\varphi) + \nonumber \\
    & \null\qquad {} + \frac{128 M q_0 d \Delta t'}{15\pi} \frac{t}{(t^2-R_c^2)^4} \left[ 2t^2 + R_c^2 - 3R_c^2\cos(2\theta) \right] \unitvec{z}, \\
    \dolate{\vec B}_{\mathrm{simple}} &\approx - \frac{128 M q_0 d \Delta t'}{15\pi} \frac{R_c(5t^2+R_c^2)}{(t^2-R_c^2)^4} \sin\theta \underbrace{(-\unitvec{x}\sin\varphi + \unitvec{y}\cos\varphi)}_{=\unitvec{\varphi}}.
\end{align}

\paragraph{Sine Pulse}

Similarly, for the sine pulse~(\ref{eq:q-simple}) we find
\begin{align}
    \dolate{\vec E}_n &\approx \frac{48 (-1)^n M q_0 d \Delta t^{\prime 2}}{n\pi^2} \frac{R_c^2(7t^2+R_c^2)}{(t^2-R_c^2)^5} \sin(2\theta) (\unitvec{x} \cos\varphi + \unitvec{y} \sin\varphi) - \\
    & \null\qquad {} - \frac{16 (-1)^n M q_0 d \Delta t^{\prime 2}}{n\pi^2} \frac{\cos\theta}{(t^2-R_c^2)^5} \left[ 10t^4 + 13 t^2 R_c^2 + R_c^4 - 3R_c^2 (7t^2+R_c^2)\cos(2\theta) \right] \unitvec{z}, \nonumber \\
    \dolate{\vec B}_n &\approx \frac{96 (-1)^n M q_0 d \Delta t^{\prime 2}}{n \pi^2} \frac{t R_c (5t^2+3R_c^2)}{(t^2-R_c^2)^5} \sin\theta \underbrace{(-\unitvec{x}\sin\varphi + \unitvec{y}\cos\varphi)}_{=\unitvec{\varphi}}.
\end{align}

\paragraph{Energy Density}

As stated at the beginning of this section, the energy and momentum densities will be calculated from the Poynting vector,
\begin{equation}
    \vec S = \vec E \times \vec B,
    \label{eq:Poynting-vector}
\end{equation}
which is just the momentum density (the $T^{0i}$ components of the  stress energy tensor).

In the above calculation we have already expanded to leading order in the size of the dipole, $d$, and the interval of the pulse, $\Delta t'$.
While the magnetic field points in the $\unitvec{\varphi}$ direction, just like  in the flat spacetime case, the electric field does not appear to point in the $\unitvec{\theta}$ direction.
This is due to not fully applying the approximations for the radiation at infinity.
(In the next section a finite observer-source separation will be discussed where this issue will be revisited.)

Recall that for this calculation we are interested in the far-field radiation so that we have the limit $R_c\rightarrow\infty$.
Like in the scalar case, the position of the center of the source (dipole), $\vec x_c'$, is fixed.
We can proceed as in section~\ref{sec:scalar-radiation-approximation}, in particular the discussion starting with eq.~(\ref{eq:Rc}) that resulted in the expressions in eqs.~(\ref{eq:scalar-T0R-simple}) and (\ref{eq:scalar-T0R-sine}), though now we will first apply the steps to the fields in the far-field limit.

\paragraph{Simple Pulse}
The approximations for a simple pulse produce
\begin{align}
    \dolate{\vec E}_{\mathrm{simple}} &\approx -\frac{16 M q_0 d \Delta t'}{5\pi r_c^4 (1-\cos\theta)^4} \left( \frac{\sin\theta}{r} \right) \unitvec{\theta}, \\
    \dolate{\vec B}_{\mathrm{simple}} &\approx -\frac{16 M q_0 d \Delta t'}{5\pi r_c^4 (1-\cos\theta)^4} \left( \frac{\sin\theta}{r} \right) \unitvec{\varphi},
\end{align}
so that
\begin{equation}
    \dolate{\vec S}_{\mathrm{simple}} \approx \left( \frac{16 M q_0 d \Delta t'}{5\pi r_c^4 (1-\cos\theta)^4}\right)^2 \left( \frac{\sin\theta}{r} \right)^2 \unitvec{r}.
\end{equation}

\paragraph{Sine Pulse}
Similarly, the approximations for the sinusoidal pulse produce
\begin{align}
    \dolate{\vec E}_n &\approx \frac{24 (-1)^{n} M q_0 d \Delta t'^2}{n \pi^2 r_c^5 (1-\cos\theta)^5} \left( \frac{\sin\theta}{r} \right) \unitvec{\theta}, \\
    \dolate{\vec B}_n &\approx \frac{24 (-1)^{n} M q_0 d \Delta t'^2}{n \pi^2 r_c^5 (1-\cos\theta)^5} \left( \frac{\sin\theta}{r} \right) \unitvec{\varphi},
\end{align}
so that
\begin{equation}
    \dolate{\vec S}_n \approx \left( \frac{24 M q_0 d \Delta t'^2}{n \pi^2 r_c^5 (1-\cos\theta)^5} \right)^2 \left( \frac{\sin\theta}{r} \right)^2  \unitvec{r}.
\end{equation}

Notice that in the far-field regime the late-time tail behavior follows the same pattern as that for the null radiation.
In particular, $|\vec E|=|\vec B|$, $\vec E$ only has a $\unitvec{\theta}$ component, $\vec B$ only has a $\unitvec{\varphi}$ component, and thus $\vec S$ is radial: momentum flows away from the source.

To calculate the energy radiated to infinity we can again follow the same procedure as in the scalar case from section~\ref{sec:scalar-radiation-approximation}.
Here we note that for the electromagnetism case $T_{0i} n^i = -\vec S\cdot\unitvec{r} = -|\vec S|$; the standard textbook expression.
We again must be careful with integrating too close to the line-of-sight.
Using the same approximations as in the scalar case we find
\begin{align}
    P^{\mathrm{tail}}_{\mathrm{simple}} &\approx -\frac{2^{15} M^2 q_0^2 d^2 \Delta t'^2 r_c'^4}{75\pi \ell^{12}}, \\
    P_n^{\mathrm{tail}} &\approx -\frac{73728 M^2 q_0^2 d^2 \Delta t'^4 r_c'^6}{n^2 \pi^3 \ell^{16}}.
\end{align}
Putting everything together, the energy of the dipole source that escapes to infinity in the tail is
\begin{align}
    \Delta E^{\mathrm{tail}}_{\mathrm{simple}} &\approx -\frac{2^{16} M^2 q_0^2 d^2 \Delta t'^3 r_c'^4}{75\pi \ell^{12}}, \\
    \Delta E_n^{\mathrm{tail}} &\approx -\frac{1}{n^2}\frac{147456 M^2 q_0^2 d^2 \Delta t'^5 r_c'^6}{ \pi^3 \ell^{16}}.
\end{align}

\paragraph{Tail/Null radiation comparison}

The null radiation in a flat spacetime that escapes to infinity can again be calculated as was done for the scalar case.
Here we find
\begin{align}
    \Delta E^{\mathrm{flat}}_{\mathrm{simple}} &\approx -\frac{64 q_0^2 d^2}{15 \pi \Delta t'^3}, \\
    \Delta E_n^{\mathrm{flat,sine}} &\approx -\frac{n^4\pi^3 q_0^2 d^2}{6 \Delta t'^3}.
\end{align}

Again using the fact that the mass distribution is required to be far from the LOS, we can compare the energy lost in the late-time tail to the energy carried by a null signal in flat spacetime.
This produces the ratios
\begin{align}
    \frac{\Delta E^{\mathrm{tail}}_{\mathrm{simple}}}{\Delta E^{\mathrm{flat}}_{\mathrm{simple}}} &\approx
    \frac{2^{10}}{5}
        \left(\frac{M}{\ell}\right)^2
        \left(\frac{\Delta t'}{\ell}\right)^6
        \left(\frac{r_c'}{\ell}\right)^4
    , \\
    \frac{\Delta E_n^{\mathrm{tail}}}{\Delta E_n^{\mathrm{flat}}} &\approx
    \frac{884736 n^6}{\pi^6}
        \left(\frac{M}{\ell}\right)^2
        \left(\frac{\Delta t'}{\ell}\right)^8
        \left(\frac{r_c'}{\ell}\right)^6
    ,
\end{align}
where $\ell$ is again a length scale associated with the size of the mass distribution.
Once again we see small ratios  ($M/\ell$, $\Delta t'/\ell$) raised to powers, balancing large numerical prefactors and a large ratio $r_c'/\ell$ raised to a power.
Using the same set of cosmological-star inspired parameters for the simple pulse as was done for the scalar radiation ratio~(\ref{eq:scalar-energy-ratio}), we find for the vector case that
\begin{equation}
    \frac{\Delta E^{\mathrm{tail}}_{\mathrm{simple}}}{\Delta E^{\mathrm{flat}}_{\mathrm{simple}}} \sim 4\sci{-58}.
\end{equation}
This is significantly smaller that in the scalar case, though the same caveats apply.
The very strong dependence on the scale $\ell$ suggests that considering more compact perturbers would enhance the tail signal.
However, we are currently limited by the weak coupling approximation $M\ll \ell$, and by the requirement that the null signal and the tail be well separated in time.

Particularly surprising to us is the strong dependence of the answer on the specifics of the source signal, with the simple pulse and the sine pulse eliciting different dependencies on the pulse frequency, the perturber size, and the source-perturber-observer geometry.

We again remind the reader that we have in this section considered the vector tail only in the presence of a point-mass perturber.
This provides only a lower limit on the amount of radiation in the tail, and our experience from the scalar case suggests that it may be a significant underestimate.
We proceed to consider a finite-size perturber.

\subsection{Vector radiation with compact perturber}
\label{subsec:vector-compactperturber}

For scalar radiation, the transition from a point-mass perturber to a finite-size perturber held important lessons (see section~\ref{subsec:scalar-compactperturber}).
We discovered that the portion of the tail radiation coming from the overlap of the constant-travel-time ellipsoid with the mass distribution --- what we referred to as the middle-time tail --- greatly exceeded the portion from the late-time tail.
Moreover, whereas the late-time tail appeared to originate from the source, the middle-time tail originated from the perturbing mass.

Here we see that these results carry over to the massless vector --- e.g.\ electromagnetic radiation.

\subsubsection{Approximations}

\begin{figure}
  \centerline{\includegraphics{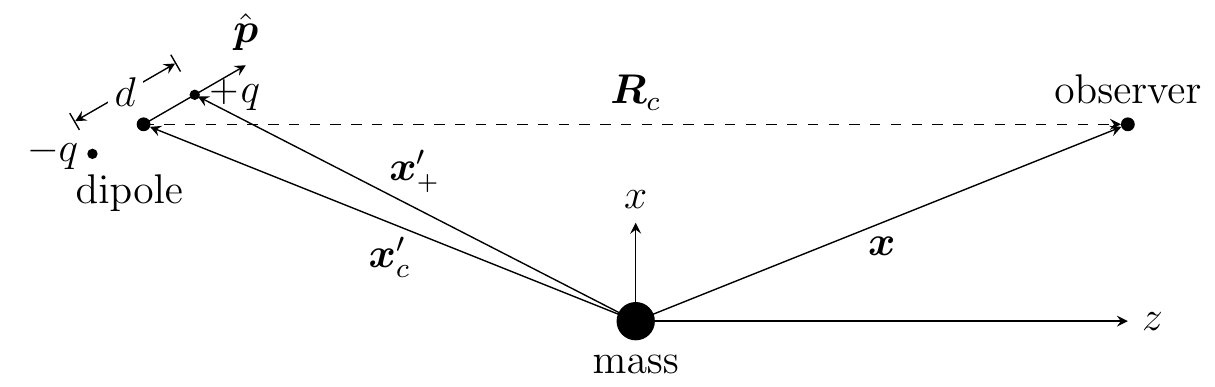}}
  \caption{Coordinate system for a finite-size mass distribution with a dipole source and an observer.
  The coordinate system has its origin at the center of the mass distribution.
  Without loss of generality the center of the dipole and center of the mass distribution lie in the $xz$-plane.
  The dipole has an arbitrary orientation (the $\unitvec{p}$ direction) but the center of the dipole and the observer lie along a line parallel to the $z$-axis of the coordinate system.
  Not drawn to scale.
  }
  \label{fig:extended_mass-coordinate-system}
\end{figure}

The calculation for massless vector radiation in the spacetime of a finite-size perturber resembles that for the scalar radiation as discussed in section~\ref{subsec:scalar-approximations}.
We again slightly modify the coordinate system used for the calculation, keeping most of the details the same, but now allow for the dipole to have an arbitrary alignment for a fixed observer location (the opposite choices were made for the study the radiation at infinity).
Without loss of generality we can choose the center of the dipole source, the center of the mass distribution, and the observer to lie in the $xz$-plane with the origin of a coordinate system at the center of the mass distribution.
Further, we can choose the center of the dipole and the observer to lie along a line parallel to the $z$-axis of this coordinate system.
Finally, we allow the dipole to point in an arbitrary direction, $\unitvec{p}$, as shown in figure~\ref{fig:extended_mass-coordinate-system}.
Recall that in this work we restrict the mass distribution to be equidistant from the center of the dipole and the observer so that $|\vec x|=|\vec x'_c|$.
For calculations it will be convenient to shift to a coordinate system with origin at the center of the dipole.
In this coordinate system the arbitrary orientation of the dipole is represented by the direction
\begin{equation}
    \unitvec{p} = (\sin\theta_d \cos\varphi_d, \sin\theta_d \sin\varphi_d, \cos\theta_d).
\end{equation}
For a dipole of size $d$ the source can be written in terms of a charge and current density; translating to the aforementioned coordinate system with origin at the center of the dipole, i.e.\ $\vec x' = \vec x'_c + r'_d\, \unitvec{p}$:
\begin{align}
    J^{0}(x') &= q(t') \delta\!\left(\frac{d}{2} - r'_d\right) \left[ \delta(\cos\theta'_d-\cos\theta_d) - \delta(\cos\theta'_d+\cos\theta_d) \right]  \delta(\varphi'-\varphi'_d) , \\
    \vec J(x') &= i(t') \Theta\!\left(\frac{d}{2} - r'_d\right) \left[ \delta(\cos\theta'_d-\cos\theta_d) - \delta(\cos\theta'_d+\cos\theta_d) \right] \delta(\varphi'-\varphi'_d) \, \unitvec{p},
\end{align}
where again $i(t')\equiv\dderiv q(t')/\dderiv t'$.

The field-strength tensor is given by
\begin{equation}
    F_{\mu\nu} = \int \dderiv^4 x' (\partial_\mu G_{\nu\beta'} - \partial_\nu G_{\mu\beta'}) J^{\beta '}.
\end{equation}
In the metric signature used in this work, the electric and magnetic field components are
\begin{equation}
    \vec E = (F_{0x}, F_{0y}, F_{0z}), \quad \vec B=(F_{zy}, F_{xz}, F_{yx})
\end{equation}

For a very small dipole, $d\ll a, R_c$, we keep only the $\mathcal{O}(d)$ contributions.
For $J^0(x')$ the delta functions lead to expanding the integrand to first order in $r'_d$, whereas in $\vec J(x')$ the integral over the step function is already of order $d$ so we expand the integrands to zeroth order in $r'_d$.
Explicitly, the electric field components are
\begin{align}
    F_{0j} &= \int \dderiv^4 x' \left[ \left( \partial_t G_{j0'} - \partial_j G_{00'} \right) J^{0}(x') + \left( \partial_t' G_{jk'} - \partial_j G_{0k'} \right) J^{k}(x') \right], \nonumber \\
    &\approx d \int \dderiv t' p^k \left\{ q(t') \partial_k \left[ \partial_t G_{j0'}(x,t',\vec x'_c) - \partial_j G_{00'}(x,t,\vec x'_c) \right] + \right. \\
    & \left. \null\qquad\qquad {} + i(t') \left[ \partial_t G_{jk'}(x,t',\vec x'_c) - \partial_j G_{0k'}(x,t',\vec x'_c) \right] \right\}, \nonumber
\end{align}
and similarly for the magnetic field components, $F_{ij}$.

The integral over $t'$ can again be approximated as in the scalar case.
For the small period pulse with some envelope $s(t'_0)$, a Taylor expansion leads to a separation of the convolution integral in a similar manner (see~(\ref{eq:field-scalar-Taylor}) for comparison).
Here we find
\begin{align}
    F_{0j} &\approx d p^k \left\{ \left( \frac{1}{m!} \int \dderiv t' q(t') t'^m \right) \right. \times \nonumber \\
    & \null\qquad\qquad\quad {}\times \int_{t'_{01}}^{t'_{02}} \dderiv t'_0 s(t'_0) \partial_k \left[ \partial_t\partial^m_{t'} G_{j0'}(x,t'_0,\vec x'_c) - \partial_j \partial^m_{t'} G_{00'}(x,t'_0,\vec x'_c) \right] +  \nonumber \\
    & \null\qquad\quad {} + \left( \frac{1}{(m+1)!} \int \dderiv t' i(t') t'^{m+1} \right) \times \nonumber \\
     & \left. \null\qquad\qquad\quad {}\times\int_{t'_{01}}^{t'_{02}} \dderiv t'_0 s(t'_0) \left[ \partial_t \partial^{m+1}_{t'} G_{jk'}(x,t'_0,\vec x'_c) - \partial_j \partial^{m+1}_{t'}G_{0k'}(x,t'_0,\vec x'_c) \right] \right\}, \\
    F_{ij} &\approx d p^k \left\{ \left( \frac{1}{m!} \int \dderiv t' q(t') t'^m \right) \right. \times \nonumber \\
    & \null\qquad\qquad\quad {}\times \int_{t'_{01}}^{t'_{02}} \dderiv t'_0 s(t'_0) \partial_k \left[ \partial_i \partial^m_{t'} G_{j0'}(x,t'_0,\vec x'_c) - \partial_j \partial^m_{t'} G_{i0'}(x,t'_0,\vec x'_c) \right] +  \nonumber \\
    & \null\qquad\quad {} + \left( \frac{1}{(m+1)!} \int \dderiv t' i(t') t'^{m+1} \right) \times \nonumber \\
     & \left. \null\qquad\qquad\quad {}\times\int_{t'_{01}}^{t'_{02}} \dderiv t'_0 s(t'_0) \left[ \partial_i \partial^{m+1}_{t'} G_{jk'}(x,t'_0,\vec x'_c) - \partial_j \partial^{m+1}_{t'}G_{ik'}(x,t'_0,\vec x'_c) \right] \right\}.
\end{align}
In these expressions we have employed the shorthand notation
\begin{equation}
    \partial^m_{t'} G_{\mu\nu'}(x,t'_0,\vec x'_c) \equiv \left. \frac{\partial^m G_{\mu\nu'}(x,t',\vec x'_c)}{\partial t'^m} \right|_{t'=t'_0}.
\end{equation}

\subsection{Results}

\begin{figure}
    \centering
    \begin{subfigure}[t]{0.475\textwidth}
        \centering
        \includegraphics[width=\textwidth]{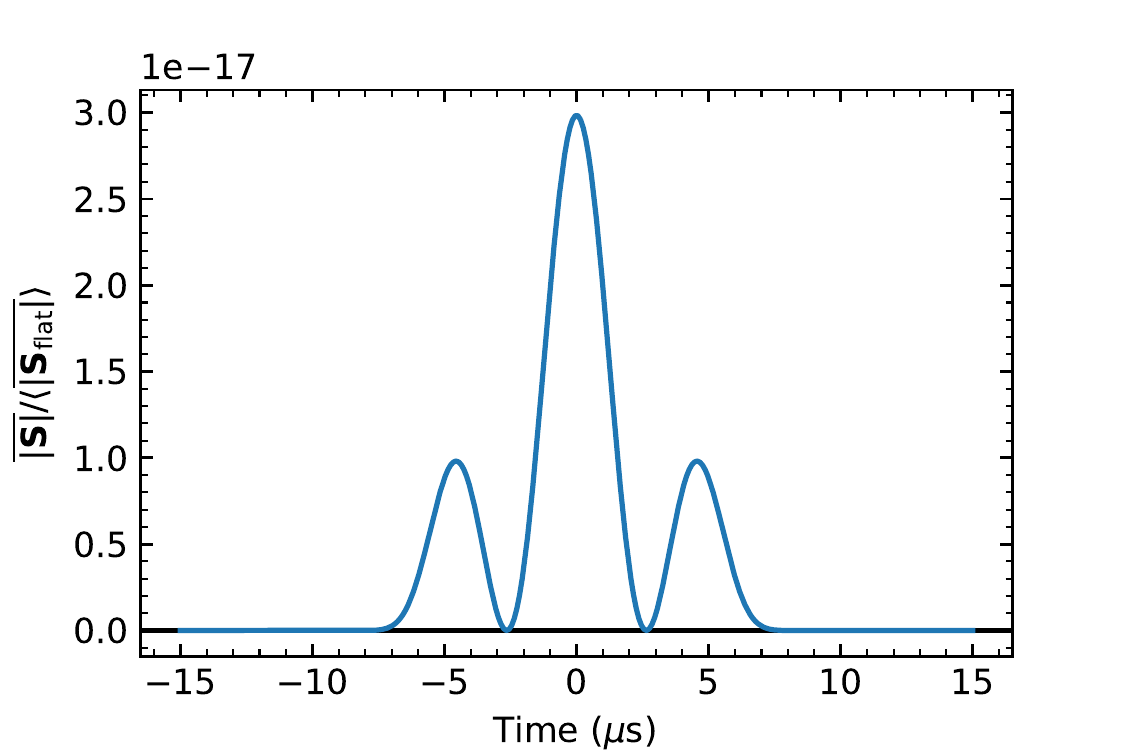}
        \caption{cosmological-star inspired parameters.}
        \label{fig:Cosmostar_Simple_100m_EM_Sratio}
    \end{subfigure}
    \hfill
    \begin{subfigure}[t]{0.475\textwidth}
        \centering
        \includegraphics[width=\textwidth]{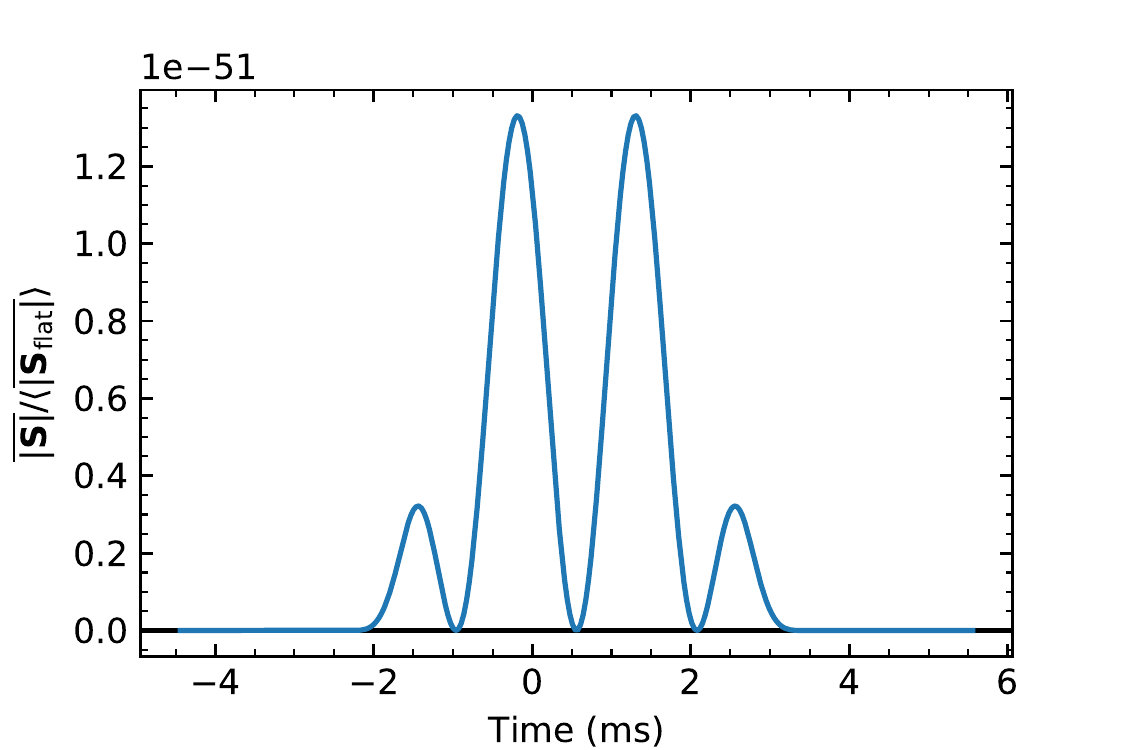}
        \caption{Lunar inspired parameters.}
        \label{fig:Lunar_Simple_100m_EM_Sratio}
    \end{subfigure}
    \vskip\baselineskip
    \caption{Energy density ratio for electromagnetic radiation from a simple pulse.
    The time on the $x$-axis is shifted as in figure~\ref{fig:CosmoStarSimple}.}
    \label{fig:Simple_100m_EM_Sratio}
\end{figure}

 \begin{figure}
    \centering
    \begin{subfigure}[t]{0.475\textwidth}
        \centering
        \includegraphics[width=\textwidth]{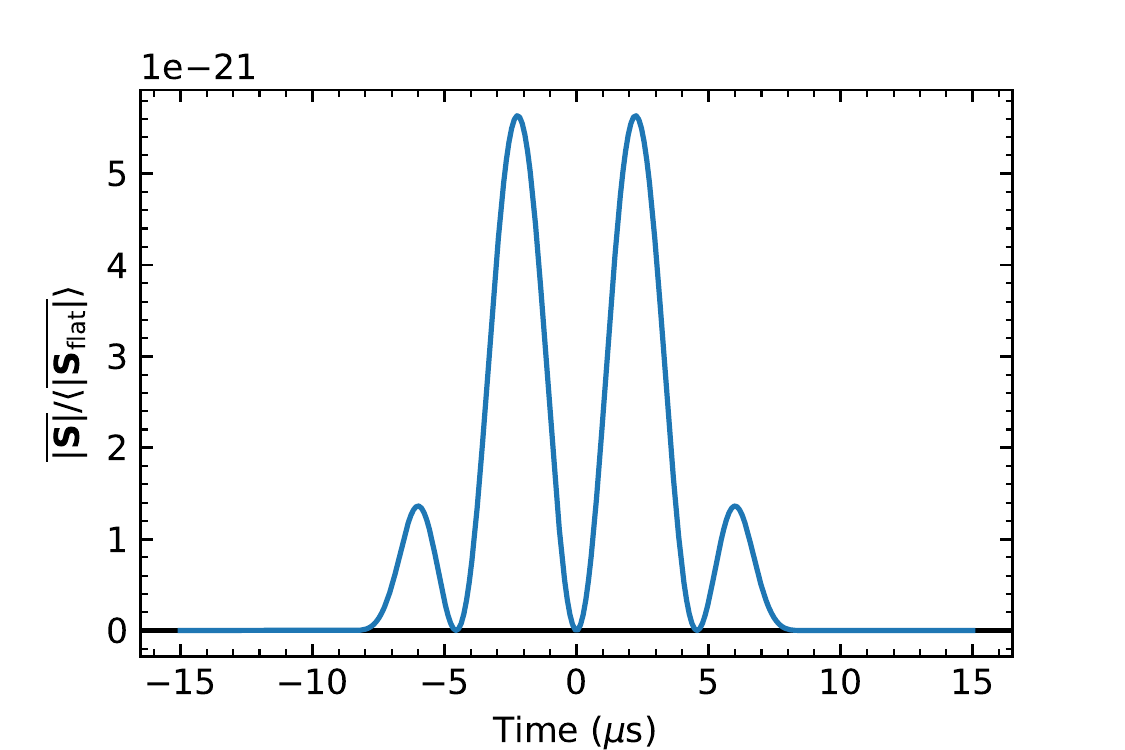}
        \caption{cosmological-star inspired parameters.}
        \label{fig:Cosmostar_Sine_100m_EM_Sratio}
    \end{subfigure}
    \hfill
    \begin{subfigure}[t]{0.475\textwidth}
        \centering
        \includegraphics[width=\textwidth]{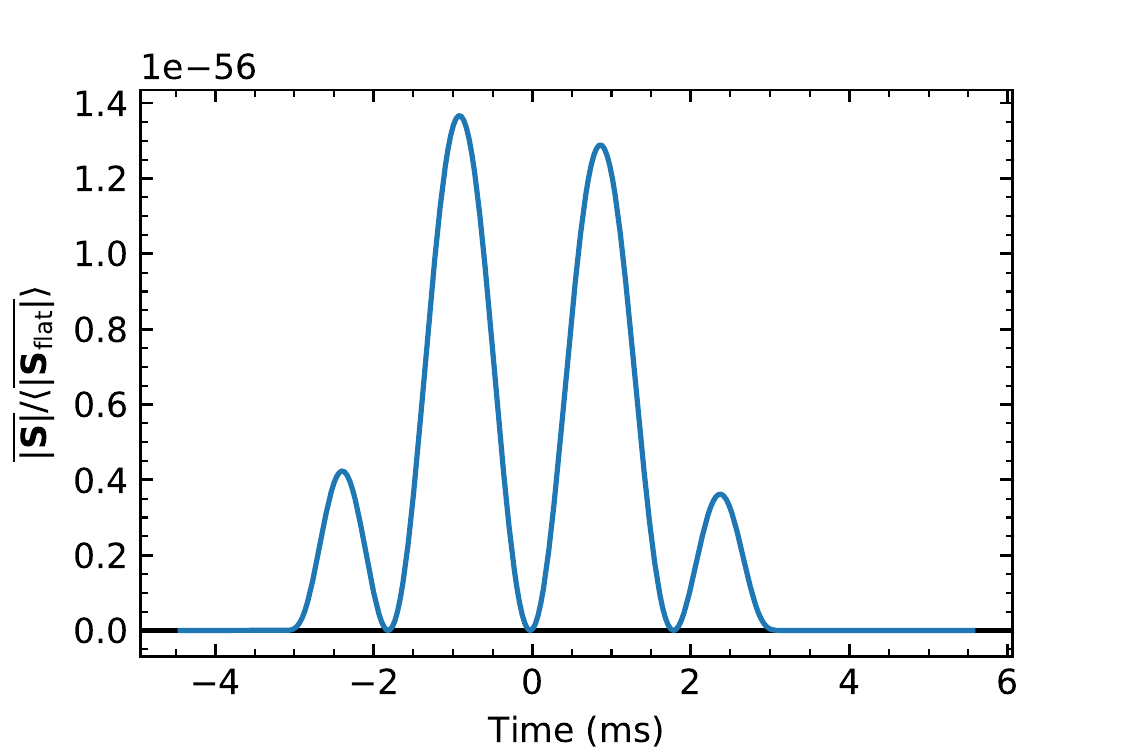}
        \caption{Lunar inspired parameters.}
        \label{fig:Lunar_Sine_100m_EM_Sratio}
    \end{subfigure}
    \vskip\baselineskip
    \caption{Energy density ratio at the observer as a function of time for electromagnetic radiation from a sine pulse. 
    The time on the $x$-axis is shifted as in figure~\ref{fig:CosmoStarSimple}.}
    \label{fig:Sine_100m_EM_Sratio}
\end{figure}

\begin{figure}[ht]
    \centering
    \includegraphics[width=\textwidth]{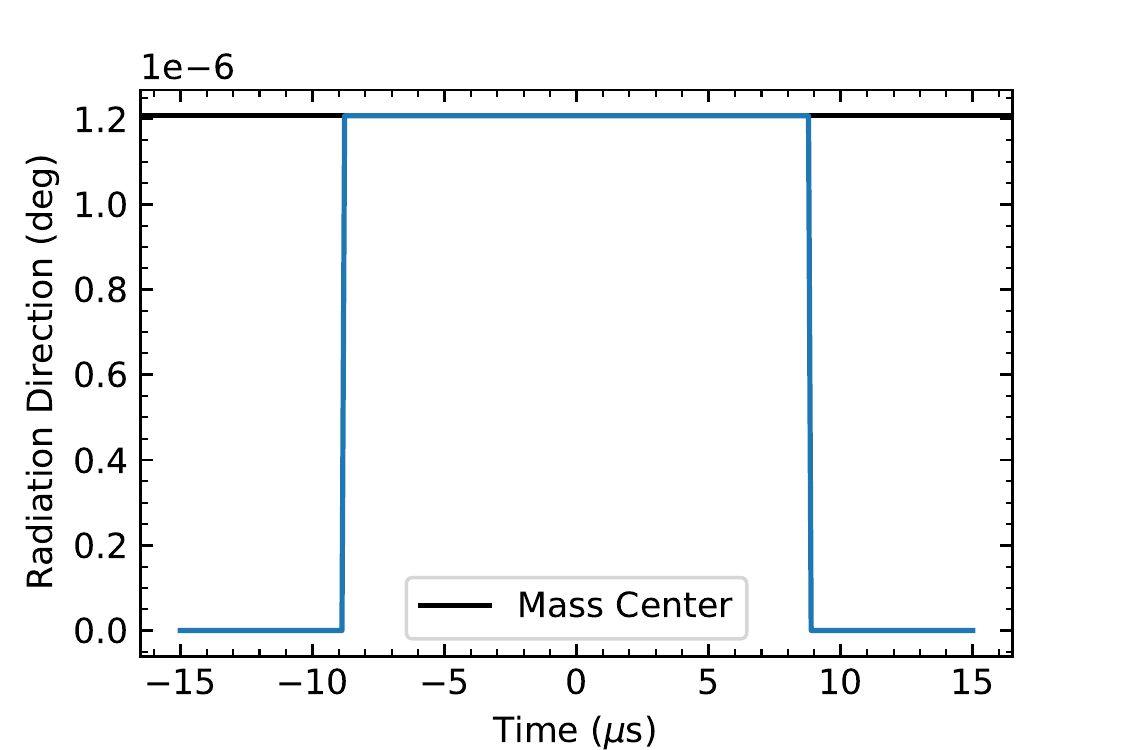}
    \caption{Direction of electromagnetic tail radiation as a function of time for the cosmological-star-inspired parameters with the simple pulse. The zero angle represents the direction to the source and the $1.2\times10^{-6}$ degree angle is the direction to the perturber center.
    The  middle-time tail is seen to come from the perturber, while the late-time tail comes from the source. 
    The time on the $x$-axis is shifted as in figure~\ref{fig:CosmoStarSimple}.}
    \label{fig:EM_radiation_direction}
\end{figure}

With the calculations above, we can now calculate the electromagnetic tail for the same set of parameters as studied in section~\ref{subsec:scalar-results} for the scalar case.
Since the electromagnetic field has more components (six instead of one for the scalar field) and since the results are going to be similar, we restrict our presentation to the two most important quantities that were calculated for the scalar field: the ratio of energy in the tail as compared to the null radiation, and the direction of the radiation.
One important difference that must be accounted for in the calculation is that the electromagnetic source, a dipole, has an orientation.
The calculations discussed above calculate the field components for a single dipole with an arbitrary orientation, $\unitvec{p}$.
For an unpolarized source we must average over all dipole orientations,
\begin{equation}
    \overline{|\vec S|} \equiv \frac{1}{4\pi} \int |\vec S|\,\dderiv\Omega_d,
\end{equation}
where $\dderiv\Omega_d=\sin\theta_d\, \dderiv\theta_d\, \dderiv\phi_d$.
For the flat spacetime we also average over the duration of a pulse,
\begin{equation}
    \langle \overline{|\vec S_{\mathrm{flat}}|} \rangle \equiv \frac{1}{2\Delta t'} \int_{-\Delta t'}^{+\Delta t'} \overline{|\vec S_{\mathrm{flat}}|}\,\dderiv t'.
\end{equation}

To understand the tail vector radiation, we first take a close look at figures \ref{fig:Simple_100m_EM_Sratio} and  \ref{fig:Sine_100m_EM_Sratio}  where we present plots of the ratio of the tail to null energy for both the cosmological-star and the lunar-inspired parameter cases.
We see more peaks and dips in the vector case, as it involves one higher derivative than the scalar, however the broad conclusions are similar:
there is no early-time tail; there is a late-time tail;
the middle-time tail is much larger than the late-time tail.

Next, the direction of the incoming vector radiation is shown in figure~\ref{fig:EM_radiation_direction} for the cosmological-star inspired parameters with the simple pulse.
(The lunar inspired parameters produce a similar result.)
Once again the fact that the middle-time tail comes from the direction of the perturber while the late-time tail comes from the source is reproduced. 

The electromagnetic case has less tail radiation than the scalar for which we have no specific physical explanation, i.e.\ beyond the mathematical result.
Nevertheless the results are not wholly discouraging for eventual observation --- the Poynting flux in the middle-time tail from the cosmological-star inspired case is $3\sci{-17}$ times that in the null signal from the source for the simple pulse as shown in figure \ref{fig:Cosmostar_Simple_100m_EM_Sratio}.
While that is small, recall that we have not yet performed an exhaustive search of the parameter space of source shapes, observational geometry, and perturber profile; and that the scalar case suggests that to maximize the tail signal we will want to go beyond some of the approximations employed in this work.

\section{Conclusions}
\label{sec:conclusions}

In this paper we have computed to leading order (in the weak field) the radiative power that fills  the interior of the lightcone compared to what travels on the surface of the lightcone.
This is known as ``tail'' radiation, and is a well known, long-considered phenomenon.
We have considered both massless scalar and massless vector (eg.\ electromagnetic) radiation.
We have done so for the first time, to our knowledge, for  finite-size, static, spherically symmetric, weak-field perturbers, for all times in which the tail radiation is non-vanishing.
This proves to be crucially different from strong-field perturbers such as black holes (or point perturbers in perturbative approaches).

The vanishing of the ``early-time'' tail (i.e.\ arriving before the time it would take to travel from source to perturber to observer at the speed of light) due to a Schwarzschild black hole is well known.
We confirm this is also the case for our finite-size  perturbers --- it is not a black-hole property.
For these perturbers, there is no tail radiation before the shortest time a signal could travel from source to a point on the perturber surface and then to the observer at speed of light.

The existence of a ``late-time'' tail (arriving after the final time for which a signal could travel from source to perturber to observer at the speed of light) is also well known for black holes.
It has also been investigated for a static, finite-size, compact, spherically symmetric , weak-field mass distribution.
It is well established that this late-time tail carries an outward flux of energy to infinity, and has even been suggested \cite{2001PhRvD..63f3003M, 2002CQGra..19..953K} that for very special source-perturber-observer geometries and long wavelength radiation this late-time tail could carry a significant fraction of the detectable energy to the observer.
This is the first time that it has been established that, oddly, this flux appears to emanate from the original source, and not from the perturber.

The existence of a middle-time tail is an important new feature of weak-field finite size perturbers.
This middle-time tail arrives during the period when a signal travelling at the speed of light could propagate from the source to some point in the perturbing mass distribution and thence to the observer.
It thus follows the (vanishing) early-time tail, and precedes (by definition) the late-time tail.
The middle-time tail carries significantly more energy than the late-time tail for the observational situations we have explored.
It also comes from the direction of the perturber.
It, and not the late-time tail, thus seems to justify the interpretation of having ``scattered'' from the metric perturbation.
The  middle-time tail also has considerable structure --- peaks and dips --- that reflect some convolution of the source pulse shape and the perturbing mass distribution.

We have confirmed these observations for both scalar radiation and electromagnetic radiation.
In a large potential parameter space, we have considered a small number of idealized examples.
One is similar to a large star at cosmological distances;
the other mimics the moon at its distance from the Earth.
We have assumed that the perturbers are transparent to the radiation (except gravitationally) --- a good assumption for certain hypothetical scalars such as axions and axion-like particles, and also for electromagnetic radiation for certain perturbers (eg.\ galaxy halos).
Nevertheless, we believe the insights gained are valuable beginnings for future investigations --- of gravitational waves, of neutrinos, of axions, of electromagnetic radiation, and for more specifically chosen observational parameters.

We are particularly encouraged  that the energy in the middle-time tail while a small fraction of what is found on the lightcone is a much larger fraction than in the late-time tail.
Large enough that one can begin to imagine an observational quest to detect it, especially if a more thorough investigation of the parameter space yields examples of still-larger fractions.
The prospect of detectable tail radiation is exciting.
We can imagine mapping the metric perturbations of the universe on a vast range of scales using the echos of bright signals scattering from them.

We have long been aware that radiation responds to the dimples of an inhomogeneous curved  spacetime by bending its path --- weak and strong lensing observations are by now a standard tool of modern astrophysics and cosmology.
But those dimples  do more than bend the null geodesics of passing null radiation.
Inspecting the massless scalar and vector Green's functions in these inhomogeneous spacetimes teaches us that unsmooth spacetime is not entirely transparent.
Inhomogeneous geometry scatters radiation, and that scattered radiation has real effects.
We had earlier 
seen its influence on the self-force of a body orbiting a central mass.
It has also been widely known to cause the interior of the forward light cone of a source to contain energy density --- in other words to induce some portion of a signal to arrive after the direct light-travel-time from source to observer as so-called tail radiation.

In this work we have presented important new features of that tail radiation.
We have shown that the late-time portion of the tail radiation generically contains a small fraction of the energy in the tail.
Most of the tail radiation carried arrives during the middle-time tail when a signal can travel from the source to the perturber and then to the observer at the speed of light.
That  middle-time tail has much more energy than the late-time tail that follows it, or than the vanishing early-time tail that precedes it.
Furthermore,  the observer sees this middle-time tail radiation pointing back to the perturber, whereas the late-time tail appears to travel down the line-of-sight, ie.\ to emanate from the source.
It is therefore crucial to resolve point perturbers into real mass distributions.

These conclusions are  valid only for static, spherically symmetric, compact, transmissive, weak-field perturbers, for short wavelength radiation, and for perturbers sufficiently far from the LOS\@.
We leave the study of what happens around the LOS for future investigation.
A wider range of observational geometries, perturber structures,  and signal sources should and will be explored.
Nevertheless, they suggest a world of opportunity to exploit the tail radiation to discover or probe a wide range of phenomena --- from mapping dark matter concentrations to discovering axions, for measuring the large scale and small scale geometry and thence  contents of the universe, for probing the interiors of neutron stars and other compact objects.

We have just begun this program.
We have presented preliminary explorations.
These results point to the need to search more broadly, to extend these techniques beyond the approximations and limits we and others have employed and to explore the tails of gravitational waves.
However there remains the  tantalizing possibility that we can use the echos and shadows of the signals we have long detected to explore more fully the contents of our universe.

\acknowledgments

GDS thanks Yizen Chu for early discussions about the prospective importance of tails. KP would like to thank Sotiria Angelitsi for pointing out some useful late time discussions of the tails in the literature.
GDS and KP are partly supported by a Department of Energy grant DE-SC0009946 to the particle astrophysics theory group at CWRU.

\bibliographystyle{JHEP}
\bibliography{2020_scalar_vector} 

\providecommand{\href}[2]{#2}\begingroup\raggedright\begin{thebibliography}{10}

\bibitem{Dyson:1920cwa}
F.~Dyson, A.~Eddington and C.~Davidson, \emph{{A Determination of the
  Deflection of Light by the Sun's Gravitational Field, from Observations Made
  at the Total Eclipse of May 29, 1919}},
  \href{https://doi.org/10.1098/rsta.1920.0009}{\emph{Phil. Trans. Roy. Soc.
  Lond. A} {\bfseries 220} (1920) 291}.

\bibitem{10.1111/j.1468-4004.2007.48439.x}
I.~Brown, \emph{{Dennis Walsh 1933–2005}},
  \href{https://doi.org/10.1111/j.1468-4004.2007.48439.x}{\emph{Astronomy \&
  Geophysics} {\bfseries 48} (2007) 4.39}.

\bibitem{PhysRevLett.13.789}
I.~I. Shapiro, \emph{Fourth test of general relativity},
  \href{https://doi.org/10.1103/PhysRevLett.13.789}{\emph{Phys. Rev. Lett.}
  {\bfseries 13} (1964) 789}.

\bibitem{hadamard1923lectures}
J.~Hadamard, \emph{Lectures on Cauchy's Problem in Linear Partial Differential
  Equations}. Yale University Press, 1923.

\bibitem{DewittBrehme1960}
B.~S. {DeWitt} and R.~W. {Brehme}, \emph{{Radiation damping in a gravitational
  field}}, \href{https://doi.org/10.1016/0003-4916(60)90030-0}{\emph{Annals of
  Physics} {\bfseries 9} (1960) 220}.

\bibitem{DeWitt:1964aa}
C.~M. DeWitt and B.~DeWitt, \emph{Falling charges},
  \href{https://doi.org/10.1103/PhysicsPhysiqueFizika.1.3}{\emph{Physics
  Physique Fizika} {\bfseries 1} (1964) 3}.

\bibitem{thorne}
K.~S. {Thorne} and S.~J. {Kovacs}, \emph{{The generation of gravitational
  waves. I. Weak-field sources.}},
  \href{https://doi.org/10.1086/153783}{\emph{Astrophys. J.} {\bfseries 200}
  (1975) 245}.

\bibitem{1993CQGra..10.2699B}
L.~{Blanchet} and G.~{Schafer}, \emph{{Gravitational wave tails and binary star
  systems}}, \href{https://doi.org/10.1088/0264-9381/10/12/026}{\emph{Classical
  and Quantum Gravity} {\bfseries 10} (1993) 2699}.

\bibitem{2000PhRvD..62h4034M}
E.~{Malec}, \emph{{Diffusion of the electromagnetic energy due to the
  backscattering off Schwarzschild geometry}},
  \href{https://doi.org/10.1103/PhysRevD.62.084034}{\emph{Phys. Rev. D}
  {\bfseries 62} (2000) 084034}
  [\href{https://arxiv.org/abs/gr-qc/0005130}{{\ttfamily gr-qc/0005130}}].

\bibitem{2001PhRvD..63f3003M}
R.~{Mankin}, T.~{Laas} and R.~{Tammelo}, \emph{{New approach to electromagnetic
  wave tails on a curved spacetime}},
  \href{https://doi.org/10.1103/PhysRevD.63.063003}{\emph{Phys. Rev. D}
  {\bfseries 63} (2001) 063003}
  [\href{https://arxiv.org/abs/gr-qc/0008047}{{\ttfamily gr-qc/0008047}}].

\bibitem{1995PhRvL..74.2414C}
E.~S.~C. {Ching}, P.~T. {Leung}, W.~M. {Suen} and K.~{Young}, \emph{{Late-Time
  Tail of Wave Propagation on Curved Spacetime}},
  \href{https://doi.org/10.1103/PhysRevLett.74.2414}{\emph{Phys. Rev. Lett.}
  {\bfseries 74} (1995) 2414}
  [\href{https://arxiv.org/abs/gr-qc/9410044}{{\ttfamily gr-qc/9410044}}].

\bibitem{MinoSasakiTanaka1997}
Y.~Mino, M.~Sasaki and T.~Tanaka, \emph{Gravitational radiation reaction to a
  particle motion}, \href{https://doi.org/10.1103/PhysRevD.55.3457}{\emph{Phys.
  Rev. D} {\bfseries 55} (1997) 3457}.

\bibitem{1997CQGra..14.1295N}
B.~C. {Nolan}, \emph{{A characterization of strong wave tails in curved
  spacetimes}},
  \href{https://doi.org/10.1088/0264-9381/14/5/030}{\emph{Classical and Quantum
  Gravity} {\bfseries 14} (1997) 1295}
  [\href{https://arxiv.org/abs/gr-qc/9703031}{{\ttfamily gr-qc/9703031}}].

\bibitem{QuinnWald1997}
T.~C. Quinn and R.~M. Wald, \emph{Axiomatic approach to electromagnetic and
  gravitational radiation reaction of particles in curved spacetime},
  \href{https://doi.org/10.1103/PhysRevD.56.3381}{\emph{Phys. Rev. D}
  {\bfseries 56} (1997) 3381}.

\bibitem{Pfenning_2002}
M.~J. {Pfenning} and E.~{Poisson}, \emph{{Scalar, electromagnetic, and
  gravitational self-forces in weakly curved spacetimes}},
  \href{https://doi.org/10.1103/PhysRevD.65.084001}{\emph{Phys. Rev. D}
  {\bfseries 65} (2002) 084001}
  [\href{https://arxiv.org/abs/gr-qc/0012057}{{\ttfamily gr-qc/0012057}}].

\bibitem{Poisson2011}
E.~{Poisson}, A.~{Pound} and I.~{Vega}, \emph{{The Motion of Point Particles in
  Curved Spacetime}}, \href{https://doi.org/10.12942/lrr-2011-7}{\emph{Living
  Reviews in Relativity} {\bfseries 14} (2011) 7}
  [\href{https://arxiv.org/abs/1102.0529}{{\ttfamily 1102.0529}}].

\bibitem{Chu:2011aa}
Y.-Z. {Chu} and G.~D. {Starkman}, \emph{{Retarded {G}reen's functions in
  perturbed spacetimes for cosmology and gravitational physics}},
  \href{https://doi.org/10.1103/PhysRevD.84.124020}{\emph{Phys. Rev. D}
  {\bfseries 84} (2011) 124020}
  [\href{https://arxiv.org/abs/1108.1825}{{\ttfamily 1108.1825}}].

\bibitem{2013PhRvD..88h4059H}
A.~I. {Harte}, \emph{{Tails of plane wave spacetimes: Wave-wave scattering in
  general relativity}},
  \href{https://doi.org/10.1103/PhysRevD.88.084059}{\emph{Phys. Rev. D}
  {\bfseries 88} (2013) 084059}
  [\href{https://arxiv.org/abs/1309.5020}{{\ttfamily 1309.5020}}].

\bibitem{Chu:2020aa}
Y.-Z. {Chu}, K.~{Pasmatsiou} and G.~D. {Starkman}, \emph{{Finite-size effects
  on the self-force}},
  \href{https://doi.org/10.1103/PhysRevD.101.104020}{\emph{Phys. Rev. D}
  {\bfseries 101} (2020) 104020}
  [\href{https://arxiv.org/abs/1910.02924}{{\ttfamily 1910.02924}}].

\bibitem{2002PhRvD..66d4008P}
E.~{Poisson}, \emph{{Radiative falloff of a scalar field in a weakly curved
  spacetime without symmetries}},
  \href{https://doi.org/10.1103/PhysRevD.66.044008}{\emph{Phys. Rev. D}
  {\bfseries 66} (2002) 044008}
  [\href{https://arxiv.org/abs/gr-qc/0205018}{{\ttfamily gr-qc/0205018}}].

\bibitem{2002CQGra..19..953K}
J.~{Karkowski}, E.~{Malec} and Z.~{Swierczynski}, \emph{{Backscattering of
  electromagnetic and gravitational waves off Schwarzschild geometry}},
  \href{https://doi.org/10.1088/0264-9381/19/5/308}{\emph{Classical and Quantum
  Gravity} {\bfseries 19} (2002) 953}
  [\href{https://arxiv.org/abs/gr-qc/0105042}{{\ttfamily gr-qc/0105042}}].

\bibitem{1974JMP....15..885P}
S.~{Persides}, \emph{{Scalar waves in the exterior of a Schwarzschild black
  hole.}}, \href{https://doi.org/10.1063/1.1666766}{\emph{Journal of
  Mathematical Physics} {\bfseries 15} (1974) 885}.

\bibitem{2003PhRvD..67f4024K}
J.~{Karkowski}, K.~{Roszkowski}, Z.~{{\'S}wierczy{\'n}ski} and E.~{Malec},
  \emph{{Waves in Schwarzschild spacetimes: How strong can imprints of the
  spacetime curvature be}},
  \href{https://doi.org/10.1103/PhysRevD.67.064024}{\emph{Phys. Rev. D}
  {\bfseries 67} (2003) 064024}
  [\href{https://arxiv.org/abs/gr-qc/0210041}{{\ttfamily gr-qc/0210041}}].

\end{thebibliography}\endgroup

\end{document}